\documentclass[aip,amsmath,amssymb,jcp,12pt
]{revtex4-1}
\usepackage[utf8]{inputenc}
\usepackage{amsfonts}
\usepackage{amsmath}
\usepackage{amssymb}
\usepackage{amsthm}
\usepackage{appendix}
\usepackage{enumerate}
\usepackage[T1]{fontenc}
\usepackage{datetime}
\usepackage{dcolumn} 
\usepackage{epigraph}
\usepackage[T1]{fontenc}
\usepackage{graphicx} 
\usepackage{hyperref}
\hypersetup{
    colorlinks=true,
    linkcolor=blue,
    filecolor=magenta,      
    urlcolor=cyan,
}
\usepackage{cleveref}
\usepackage{mathptmx}
\usepackage{xcolor}
\usepackage{yfonts}

\newcommand{\bfr}{{\bf r}}

\newcommand{\R}{\mathbb{R}}
\newcommand{\erf}{\mathrm{erf}}
\newcommand{\erfc}{\mathrm{erfc}}
\newcommand{\lm}{(\lambda,\mu)}

\newcommand{\bra}[1]{\ensuremath{\langle #1 \vert}}
\newcommand{\ket}[1]{\ensuremath{\vert #1  \rangle}}

\begin{document}


\title{Models and corrections: range separation for electronic interaction - lessons from density functional theory}

\author{Andreas Savin}
\affiliation{Laboratoire de Chimie Th\'eorique, CNRS and Sorbonne University \\ 4 place Jussieu, 75252 Paris cedex 05, France}
\email{andreas.savin@lct.jussieu.fr}
\begin{flushright}
\begin{minipage}{0.8\textwidth}
\noindent
The following article has been accepted by {\em The Journal of Chemical Physics}.
After it is published, it will be found at \\
\url{https://aip.scitation.org/doi/10.1063/5.0028060} 
\end{minipage} 
\end{flushright}

\begin{abstract}
Model Hamiltonians with long-range interaction yield energies that are corrected taking into account the universal behavior
 of the electron-electron interaction at short range.

Although the intention of the paper is to explore the foundations of using density functionals combined with range separation,
 the approximations presented can be used without them, as illustrated by a calculation on Harmonium.
In the regime when the model system approaches the Coulomb system, they allow the calculation of ground states, excited states, and properties,
 without making use of the Hohenberg-Kohn theorem.
Asymptotically, the technique is improvable, allows for error estimates that can validate the results. 

Some considerations for correcting the errors of finite basis sets in this spirit are also presented.

Being related to the present understanding of density functional  
 approximations, the results are comparable to those obtained with the latter, as long as these are accurate.

\end{abstract}


\maketitle

\tableofcontents
\newpage

\centerline{\today \hspace{2pt} at \currenttime}

\section{Introduction}

\subsection{Approach}

\subsubsection*{A model and its correction}

Combining methods is often encountered in electronic structure calculations. 
For example, it is present in hybrid methods for density functional approximations (DFAs), see, e.g., Ref.~\cite{Bec-JCP-93a}, or 
 when extrapolating to the complete basis set limit after coupled cluster calculations, see, e.g., Ref.~\cite{HelKloKocNog-97}.

This paper presents a model Hamiltonian.
It uses an interaction between electrons that is not the physical, Coulombic one,
 but recovers its behavior in the long-range, while having no singularity when the distance between electrons vanishes.
A parameter, $\mu$ is used to characterize the model.
In the limit $\mu=0$ the model corresponds to a system without interaction between electrons, 
 while as $\mu \rightarrow \infty$ the Coulombic interaction is reached.
For each of the models, i.e., for each $\mu$, corrections are constructed by exploiting the analytical behavior of
 the solution of the Schr\"odinger equation for the model as the Coulombic system is approached.
The method uses ``range-separation''.

While with Jastrow factors, or in F12-methods (see, e.g., Refs. \onlinecite{TewKlo-MP-10,TenNog-12}), the wave function is corrected, 
 here a modified Hamiltonian is considered to provide reliable eigenvalues, and corrections are applied to the latter.

\subsubsection*{Universality}
A reason for using a range-separated model lies in the  possibility to introduce features that are
 independent of the specific system studied (are ``universal'').
When electrons get close (in the ``short-range'') the repulsion between electrons dominates, 
 making the interaction with the nuclei (specifying the system) irrelevant for the correction. 
This motivates the exclusion of the short-range from the model Hamiltonian.

\subsubsection*{Systematic improvement}

Often one speaks about ``systematic improvement''.
However, it can have different meanings.
\begin{enumerate}
 \item The lowest level is statistical: repeated applications show that method $A$ is superior to method $B$. 
       This meaning is found in literature, e. g.,  as showing that the generalized gradient approximations (GGAs) are systematically
       improving over the local density approximation (LDA).
 \item The next level is nicely summarized by Jerome K. Percus' (NYU, New York, USA) statement~\footnote{private conversation}
       \begin{quote}
         Every time I develop a model, I like to know what the next step could be, even if it is too complicated for me to pursue it.
       \end{quote}
       It is often used in quantum chemistry, e. g., when saying that coupled cluster methods are systematically improvable, 
       as one can enlarge the one-particle basis set, or the degree of excitation from the reference configuration.
 \item A higher level is producing a bound, as, e. g., for the energy in a variational method.
       We know that the error gets smaller after the improvement, but we do not know how large the error is.
 \item An even higher level is provided by a model where error bounds can be improved. 
       Wolfgang Hackbusch (Max Planck Institute for Mathematics in the Sciences, Leipzig, Germany) once asked me~\footnote{private conversation}
       \begin{quote}
        Why do you speak about approximations?
        The word approximation comes from proximity,
         you have to know how far you are from the desired result.
        This is the meaning we use in mathematics.
        You should use another word, for example, model.
       \end{quote}
\end{enumerate}
 
This paper presents an approach going into the direction of overcoming the first level above,
 using ideas that originated from the understanding of density functional approximations.


\subsection{ Warnings}


\subsubsection*{No mathematical rigor}

The paper contains mathematical derivations.
However, a mathematician will not consider the derivations in this paper acceptable.
For those looking for rigor, everything presented in this paper can be resumed to an attempt to correct the error
 made by a model by adding a linear combination of functions selected to vanish when the model becomes exact.
The expansion coefficients are determined by giving some information about the way the model evolves toward the Coulomb system.

\subsubsection*{Asymptotic improvement}

The method presented works in an asymptotic regime (as the model approaches the Coulomb system, $\mu \rightarrow \infty$).

Fortunately, the numerical results presented here are encouraging in the sense that the method
 seems to work even when quite far from the Coulomb system ($\mu \approx 1$).

\subsubsection*{No single-determinant wave functions}
Although single determinant wave functions are present in many density functional methods, the method discussed in this paper is not restricted
 to single-determinant wave functions, except when the model corresponds to a non-interacting system ($\mu=0$). 
On one hand, this makes it more expensive than those frequently (but not always) present in range-separated application with DFAs.
On the other hand, it gives the model more flexibility, and allows the correction to be simpler. 
In the limiting case of the model becoming exact ($\mu =\infty$), the correction vanishes. 
Of course, in this limiting case, the exact Schr\"odinger equation has to be solved.

The convergence to the $N$-particle basis set limit is faster when the singularity in the Hamiltonian is eliminated,
 making the calculation of the model system faster than for that with Coulomb interaction.
 
There was also another source of interest for this kind of treatment: the deception about the treatment of static correlation
 with correlation energy density functionals~\cite{SavStoPre-TCA-86}.
Static correlation is related to degeneracy, and degeneracy is due to the external potential, and thus not a universal feature.~\cite{Sav-08}
This makes it difficult to have it described by a universal functional.
Hence the idea of going beyond the single Slater determinant treatment (as in the Kohn-Sham method) and combining
density functionals with more elaborate wave functions. 
One of the ideas was to use range-separation.~\cite{StoSav-INC-85}
 
\subsubsection*{Limited treatment of simplified wave function methods}
With the exception of basis set limitations, insufficiencies in the wave function treatment are not discussed here.
Thus, no comparison is made with those coming from correcting second order perturbation theory 
 only, e.g., Refs. \onlinecite{GolWerSto-PCCP-05, AngGerSavTou-PRA-05}, the random phase approximation, e.g., Ref. \onlinecite{TouGerJanSavAng-PRL-09},
 with coupled cluster, e.g., Ref. \onlinecite{GolWerSto-PCCP-05},
 multi-configuration self-consistent field calculations, e.g., Ref. \onlinecite{FroTouJen-JCP-07},
 multi-reference configuration interaction, e.g., Ref. \onlinecite{LeiStoWerSav-CPL-97},
 projected Hartree-Fock~\cite{GarJimScu-JCP-13}, constrained pairing mean-field theory~\cite{TsuScuSav-JCP-10},
 density matrix renormalization group~\cite{HedKneKieJenRei-JCP-15},
 fixing the nodes in diffusion Monte Carlo~\cite{SceGinBenLoo-20}, or using a specialized ansatz for describing the long range part 
 with applications such as dispersion interaction~\cite{KamTsuHir-JCP-02}.
Of course, any method that is pushed far enough to reproduce full configuration interaction to the desired accuracy, 
 like selective configuration interaction, e.g., Ref. \onlinecite{GinPraFerAssSavTou-JCP-18}, is not excluded by the present treatment. 
 
\subsubsection*{The external potential as the one-particle potential}
It is elementary knowledge that a mean-field Hamiltonian (such as Hartree-Fock or Kohn-Sham) is a better starting point
 for perturbation theory calculations than the bare external potential.
Nevertheless, for the sake of the simplicity of the argumentation, the latter is used.
The effect of relaxing this restriction will be succinctly addressed.

\subsubsection*{No use of DFAs}
This paper does not intend to present methods that correct the model using a DFA, but only to justify them.

A reason for avoiding here to use  DFAs is the need of choosing one among an amazing range of possibilities.
By what criterion should the density functional be selected for a perspective paper?
Furthermore, this paper is not a review paper, but only aims to present a viewpoint, and hopefully opens ways to improve DFAs.
 
The theoretical foundations of correcting the model used in this paper with a density functional has been established long ago 
 \cite{SavFla-IJQC-95, Sav-INC-96}: it simply results from the Hohenberg-Kohn theorem.
If the exact density is known, accurate corrections can be even constructed (even if it is exceedingly difficult, see, e.g., Refs.
 \onlinecite{PolColLeiStoWerSav-IJQC-03, TeaCorHel-JCP-10b}). 


By no means should this paper reduce the importance of using range separation with DFAs, as proven by the many successful applications, 
\cite{IikTsuYanHir-JCP-01, YanTewHan-CPL-04, VydScu-JCP-06, PeaCohToz-PCCP-06, ChaHea-JCP-08, KroSteRefBae-JCTC-12, KroKum-18} to cite just a few.
A review paper is Ref. \onlinecite{TsuHir-14}.
The reasoning below relies heavily on the experience obtained within density functional theory (DFT). 
Also, a connection to pair density functional theories (such as Refs. \onlinecite{GusMalLin-MP-04, ManCarLuoMaOlsTruGag-JCTC-14, HapPerGri-JPCL-20})
 should also be evident.
 
The applicability of asymptotic considerations not only to ground states, but also to excited states, and properties, in general,
 gives a further reason not to use the Hohenberg-Kohn theorem in this paper.

\subsubsection*{Other approaches}

This paper does not discuss closely related papers, such as presented in Refs. \onlinecite{SeiPerKur-PRA-00,AouGatRei-20, Bur-20}, or relating
 short-range DFAs with a treatment using Jastrow factors~\cite{ColSal-75, ColSal-79, FlaSav-PRA-94, SceGinBenLoo-20}.

Furthermore, uses of range separation to deal with subsystems, see, e.g., \onlinecite{QueKum-JCP-15, MosJonBrRatSch-19}, are also not treated.  

This paper does not describe methods where correlation effects are included into the construction of a model Hamiltonian.
Such methods exist for a long time and are successful.~\cite{Dav-70, BylKle-PRB-90, Pan-95, HeyScuErn-JCP-03, KruVydIzmScu-06, 
 HenIzmScuSav-JCP-07, SonTokSatWatHir-JCP-07}
They can be seen as using pseudopotentials for the electron-electron interaction, with the intention to directly answer the physical problem 
 while staying with single-determinant wave functions.
While in the pseudopotential approach the difficulty lies in defining a good model, paying it with no, or only a simple correction.
In the approach presented here, the model is to a certain degree arbitrary (and thus adaptable to requirements of the system studied), 
 all the difficulty being that of finding the right correction.

\subsubsection*{There are not enough examples in this paper}
Using the asymptotic behavior does not tell us how far we can get with them. 
Here, we use a single example, Harmonium with two electrons, for which we have an analytic solution for the ground state.
For the models where we don't have analytical solutions, we can still reduce the problem to a problem in one dimension that can be 
 accurately solved numerically.
Approximations can be followed  step by step.

Of course, this is a poor basis for generalizing the results, being far below the present standard of testing approximations on large sets 
 of systems.
Other systems treated in a similar fashion can be found, e.g., in Refs. \onlinecite{Sav-JCP-11, RebTouTeaHelSav-JCP-14, SenHedAlaKneFro-MP-16,
 AlaKneFro-16, AlaDeuKneFro-17}.

\subsubsection*{This paper presents a personal viewpoint} 
Mostly own papers are cited (in order to complement the argumentation), and many important papers (even some that produced a breakthrough)
are unfortunately not cited. The author begs to be excused for these omissions.

\subsection{Structure of the paper}
First, the model Hamiltonian used in this paper is described in more detail.
Next, the asymptotic treatment is introduced.
It is explained how it can provide corrections for the eigenvalues of the Schr\"odinger equation, and how 
 properties can be corrected.
Although local approximations  (like those used in DFAs) are not made here, some connections to them are mentioned. 
For practical reasons, it is useful to treat also the case when the models are not solved exactly, due to basis set limitations.
Their effect is compared to that of having a long-range interaction operator. 
In order to illustrate the method, some numerical results are presented for the Harmonium system with two electrons.
The paper ends with some thoughts about future developments.

\section{Range separation}

\subsection{Context}

\subsubsection*{Ewald approach}
\begin{figure}[htb]
 \begin{center}
   \includegraphics[width=0.95\textwidth]{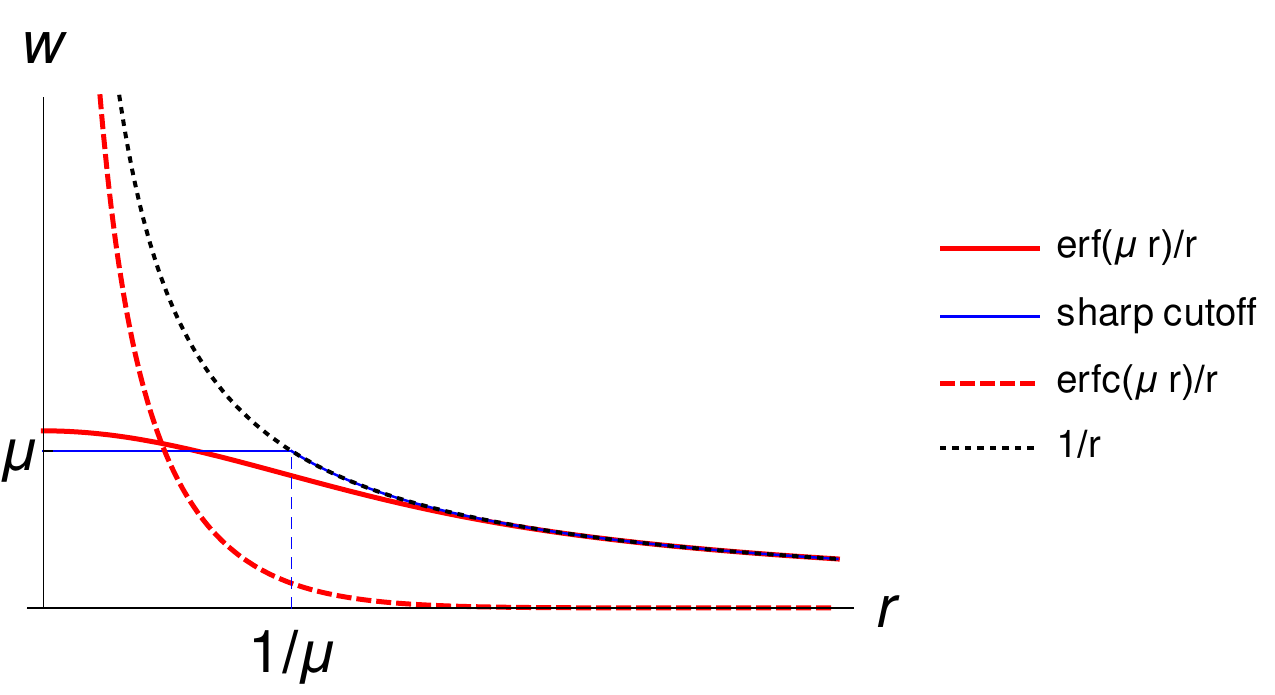}
 \end{center}
 \caption{Ewald decomposition, Eq.~\eqref{ewa-position}, of the Coulomb interaction (dotted, black) into a long range part, 
          as used in the model (full, red), 
          and a short range part (dashed, red); a rough approximation to the long-range part (full, thin, blue) is given by a function that is constant
          ($\mu$) for distances smaller than $1/\mu$, and is identical to the Coulomb interaction for large $\mu$.
          }
 \label{fig:w}
\end{figure}

In order to describe the electrostatic potential on a crystal lattice, Ewald \cite{Ewa-AP-21} found it useful to partition the Coulomb interaction
 between two particles into two parts:
\begin{equation}
 \label{ewa-position}
 \frac{1}{r} = \frac{\erf(\mu r)}{r} + \frac{\erfc(\mu r)}{r}
\end{equation}
where $r$ is the distance between particles.
A schematic plot is shown in Fig.~\ref{fig:w}.
For given $\mu$, the first term on the right-hand side of Eq.~\eqref{ewa-position} follows the Coulomb interaction when $r$ is large 
 (it is long-ranged), while the second term on the right-hand side is short-ranged.
$\mu$ has the dimensions of an inverse distance.
When $\mu=0$, the long-range part vanishes, while it approaches the Coulomb interaction when $\mu=\infty$.
The long-range part is reaching smoothly the Coulomb interaction for $r \approx 1/\mu$, and is close to $\mu$ between the origin and this point.

The long-range part in position space is short-ranged in the momentum space; the Fourier transforms of the terms in Eq.~\eqref{ewa-position} gives:
\begin{equation}
 \label{ewa-momentum}
 \frac{4 \pi}{k^2} =\frac{4 \pi}{k^2} e^{-\frac{k^2}{4 \mu^2}} + \frac{4 \pi}{k^2} \left( 1 - e^{-\frac{k^2}{4 \mu^2}} \right)
\end{equation}
From now on, when speaking about long or short-range, we refer to the components in position space, Eq.~\eqref{ewa-position}.

The choice of the separation parameter $\mu$ is a matter of compromise between the treatments.
It is analogous to the choice of basis sets: although it can be systematically improved, most often it is experience that decides its choice.

\subsubsection*{Range separation can be used to simplify the numerical treatment}
In general, short-range is more convenient for calculations (think of order $N$ calculations). 
Ewald proposed to stay in position space when the interaction is short-ranged in it, but
 to switch to momentum space for the part that is of long-ranged in position space (and of short-ranged in momentum space).

Another example of using range separation to combine methods was provided by Nozi\`eres and Pines.~\cite{NozPin-PR-58}
They noticed that the second-order perturbation theory diverges for the uniform electron gas when
 the Coulomb interaction is used, but not when a short-range interaction is used.
They thus proposed to calculate the correlation energy using the second order perturbation theory only for the short-range, while
 using another method (that has problems at short-range, the random phase approximation) for the long-range.
 
There are two unpublished papers, written in 1977 by W. Kohn (at the time at the University of San Diego, USA) and 
 W. Hanke (at the time at the Max-Planck-Institute for Solid State Physics, Stuttgart, Germany) where density functionals were used only for 
 describing the short-range interaction between electrons.
 
\subsection{Model}

\subsubsection*{Model interaction} 
In the models treated in this paper, the interaction between two electrons, one in $\bfr_1$, the other in $\bfr_2$ 
 is given by the long-range part in Eq.~\eqref{ewa-position}):
\begin{equation}
 \label{w}
 w(r_{12},\mu) = \frac{\erf(\mu r_{12})}{r_{12}} 
\end{equation}
where $r_{12}=|\bfr_1 - \bfr_2|$.
For $N$ particles, we have the interaction operator,
\begin{equation}
 \label{W}
 W(\mu)= \sum_{i<j}^N w(r_{ij},\mu)
\end{equation}

\subsubsection*{Model Hamiltonian} 
The model Hamiltonian is
\begin{equation}
 \label{H}
 H(\mu)= T + V + W(\mu) .
\end{equation}
Here $T$ is the kinetic energy operator, and $V$ the one-particle potential.
Please note that in density functional theory (DFT), one does not choose $V$ as the external potential, but adds a correction to it, 
such that for all model system (all $\mu$) the density is identical to the exact one.
Thus, in a density functional context, $V$ depends on $\mu$, is the Kohn-Sham potential for $\mu=0$, 
 and becomes the external potential only for $\mu=\infty$. 
This is useful, but there are three reasons for not doing it in this paper.
First, DFT requires the knowledge of the exact density that we do not have a priori.
Second, there is no clear criterion to choose a DFA from the multitude of the existing ones.
Third, below derivatives with respect to $\mu$ will be used. 
Once $V(\mu)$ is chosen, it is possible to use them without much difficulty, but this makes the formulas less transparent.
Thus, in this paper, $V$ is the same for all models (all $\mu$s), namely the external potential.

The Schr\"odinger equation for the model Hamiltonian is:
\begin{equation}
 \label{SE}
 H(\mu) \Psi(\mu) = E(\mu) \Psi(\mu)
\end{equation}

When no dependence on $\mu$ is explicitly given, the physical quantities are supposed: 
\begin{align}
 H & = H(\mu=\infty) \nonumber \\
 E & = E(\mu=\infty) \nonumber \\
 \Psi & = \Psi(\mu=\infty) \nonumber \\
\end{align}

\subsubsection*{Motivations for the choice of the model system}

Kato~\cite{Kat-CPAM-57} has noticed that the singularity of the Coulomb interaction showing up when the 
 distance between electrons vanishes, $r_{12} \rightarrow 0$, makes the wave function behave in a way independent of the external potential.
This can be understood physically:  when two electrons get close, the repulsion dominates, and the specific features of the external potential
 do not matter.
This behavior is ``universal''.
Constructing a model system that eliminates the singularity that appears for the Coulomb interaction at the origin  improves the convergence 
 with increasing basis set size (for wave functions written as linear combinations of Slater determinants).
Thus, the model interaction in Eq.~\eqref{w} has the double advantage: as in DFT, it eliminates the need of describing some universal 
 properties for each of the systems calculated, and it makes the latter calculation simpler. 





Of course, the use of the error function to make the separation between the ranges, Eq.~\eqref{w}, is decided by convenience. 
One can think of other functions, like the exponential (as in the Yukawa interaction \cite{SavFla-IJQC-95}, or more complex forms 
 (see, e.g., Refs. \onlinecite{LloNeeCon-15, GonAyeKarSav-TCA-16}, to have a non-singular interaction that 
 reproduces on average the interaction between electrons).
A reason to use the error function is the simplicity to implement it, as the Fourier transform, Eq.~\eqref{ewa-momentum},
 used in the calculation of two-electron integrals introduces only Gaussian factors, easy to deal with both in Gaussian and in plane wave codes.

\section{Correcting the model}

\subsection{Asymptotic regime}

\subsubsection*{Approaching the physical interaction}

Our aim is to construct corrections that become exact when the model system approaches the physical one.
There are several paths to approach it. 
For example, using our model interaction, Eq.~\eqref{w}, one can increase $\mu$.
Another way is to construct a Hamiltonian depending also on another parameter $\lambda$ that, as in conventional perturbation theory,
 reaches the physical system for $\lambda=1$,~\cite{KohMeiMak-PRL-98, SirKin-JCP-02, SirKin-IJQC-03}
\begin{equation}
 \label{H-lambda-mu}
H\lm = H(\mu) + \lambda \left( H  - H(\mu) \right)
\end{equation}
For the choice of $H(\mu)$ of Eq.~\eqref{H}, the correcting term is
 $\lambda \left( H  - H(\mu) \right)=\lambda \bar{W}(\mu)$
 where
\begin{equation}
 \label{W-bar}
  \bar{W}(\mu)=\sum_{i<j} \bar{w}(r_{ij},\mu)
\end{equation}
and
\begin{equation}
  \label{w-bar}
  \bar{w}(r,\mu)=1/r -w(r,\mu)= \frac{\mathrm{erfc}(\mu r)}{r}
\end{equation}
Note that the two-particle operator in $H(\lambda,\mu)$ can also be written as
\begin{equation}
 w(r,\mu)+\lambda \, \bar{w}(r,\mu)  = (1-\lambda) \, w(r,\mu) + \frac{\lambda}{r} 
 \label{w-with-cusp} 
\end{equation}
showing that a non-zero $\lambda$ introduces a weakened Coulomb singularity in the potential, even when $w$ is non-singular.
For this reason, in this paper, although we consider $\lambda \ne 0$ when constructing corrections, we always choose $\lambda=0$ for the model system. 
However, the formulas are easy to generalize to $\lambda \ne 0$.

The Schr\"odinger equation with $H(\lambda,\mu)$ has eigenvalues $E(\lambda,\mu)$ and eigenfunctions $\Psi(\lambda,\mu)$.
Note that the Coulomb system can be equally reached either by $\lambda=1$, or by $\mu=\infty$, 
\begin{align}
 H(\lambda=1,\mu)    & = H (\lambda, \mu = \infty)  = H \nonumber \\
 E(\lambda=1,\mu)    & = E (\lambda, \mu = \infty)  = E \nonumber \\
 \Psi(\lambda=1,\mu) & = \Psi (\lambda, \mu = \infty)  = \Psi \nonumber 
\end{align}
and that
\begin{align}
 H(\lambda=0,\mu)    & = H (\mu)\nonumber \\
 E(\lambda=0,\mu)    & = E (\mu) \nonumber \\
 \Psi(\lambda=0,\mu) & = \Psi (\mu) \nonumber 
\end{align}

We use also the notation
\begin{align}
 \bar{E}\lm & = E - E\lm \nonumber \\
 \bar{E}(\mu) & = E - E(\mu)
 \label{E-bar}
\end{align}
for the corrections to the models.

\subsubsection*{The asymptotic behavior of $\Psi(\mu)$}

When $W$ is close to the Coulomb interaction ($\mu \rightarrow \infty$), and the distance between electrons 1 and 2, $r_{12}$, is small,
the interaction of the electrons with the external potential loses its importance in the Schr\"odinger equation.
By changing variables to have an explicit dependence on $r_{12}$, we have to consider the following differential equation:
\begin{widetext}
\begin{equation}
 \left( -\partial_{r_{12}}^2 +\frac{2}{r_{12}} \partial_{r_{12}} + w(r_{12},\mu) + \lambda \, \bar{w}(r_{12},\mu) \right) \Psi(r_{12},\dots,\mu)=0
 \label{srSE}
\end{equation} 
\end{widetext}
that is valid for $r_{12} \rightarrow 0$ and $\mu \rightarrow \infty$.
After changing variables $r_{12} \rightarrow \mu r_{12}$, and staying to first order in $1/\mu$, one obtains as solution of the differential
 equation 
\begin{equation}
\label{psi-asy}
 \Psi(\dots, r_{12}, \dots; \lambda,\mu) \rightarrow c \, \phi(r_{12},\lambda,\mu) \;\;\;  \mathrm{for} \; \mu \rightarrow \infty, 
  r_{12} \rightarrow 0
\end{equation}
 where 
\begin{widetext}
\begin{equation}
 \label{phi}
 \begin{split}
 \phi(r_{12},\lambda,\mu) = 1+ \frac{\lambda}{2} r_{12} 
  + (1-\lambda) \left( \frac{1}{2 \sqrt{\pi} \mu} e^{-\mu^2 r_{12}^2} + \frac{1}{2} r_{12} \;  \erf(\mu r_{12}) 
  + \frac{1}{4 \mu^2} \frac{1}{r_{12}} \erf(\mu r_{12}) \right),  \\
   \; \mathrm{for} \; \ r_{12} \rightarrow 0, \mu \rightarrow \infty 
 \end{split}
\end{equation}
\end{widetext}
$c$ does neither depend on $r_{12}$ (short range expansion), nor on the model parameters $\lm$ (the model system is close to
 the Coulomb system), but on all the other coordinates, and the Coulomb system under consideration (that we have ``perturbed'').
The details of the calculation are too lengthy to be presented here, but the technique used is identical to that of Ref. \onlinecite{GorSav-PRA-06}.

\begin{figure}[htb]
 \begin{center}
   \includegraphics[width=0.95\textwidth]{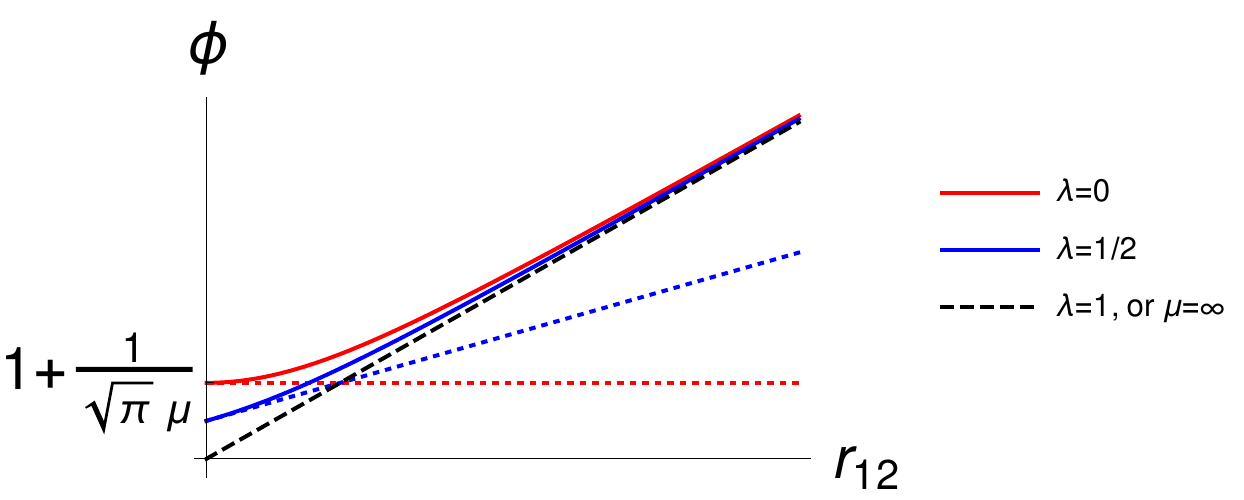}
 \end{center}
 \caption{Schematic behavior of the wave function dependence on the inter-particle distance $r_{12}$, for small $r_{12}$ and large $\mu$, 
  Eq.~\eqref{phi}: $\lambda=0$, full red  curve; $\lambda=1/2$, full blue curve; the tangents at $r_{12}=0$ are shown as dotted lines; 
  $\lambda=1$, i.e., Coulomb interaction, dashed black line.
  }
 \label{fig:phi}
\end{figure}

By expanding $\phi$ around $r_{12}=0$, we see how the cusp is approached:
\begin{widetext}
\begin{equation}
 \label{Kato-phi}
 \phi(r_{12},\lambda,\mu)= 1+\frac{1-\lambda}{\sqrt{\pi} \mu} + \frac{\lambda}{2} r_{12} + \mathcal{O}(r_{12}^2)
 ,  \; \mathrm{for} \;  \mu \rightarrow \infty 
\end{equation}
\end{widetext}
We note that the Kato behavior, $1+r_{12}/2$ is recovered for $\lambda=1$,
 and that the cusp disappears at $\lambda=0$, i.e., for the model Hamiltonian $H(\mu)$, Eq.~\eqref{H}.
The schematic behavior of $\phi$ as a function of $r_{12}$ is shown in Fig.~\ref{fig:phi}.

 
Eq.~\eqref{phi} is valid for singlet electron pairs.
For triplet pairs an analogous treatment is possible, cf. Ref. \onlinecite{GorSav-PRA-06}.
Triplet pairs contribute to higher order in $\mu$ than considered in the theoretical part of this paper.
Furthermore, as the only numerical example is a system of two electrons in a singlet state, triplet pairs are not discussed.

\subsubsection*{Determining the system-specific prefactor}

Additional information is needed about the constant $c$ that shows up in Eq.~\eqref{psi-asy}.
As $\phi$ is valid only for the small $r_{12}$, and behaves like $1+r_{12}/2$ for large $r_{12}$, normalization cannot help.~\footnote{$\phi$, 
 as given in Eq.~\eqref{phi}, is not normalizable.} 
We will not need $c$ as such because expectation values over short range operators will be used below.
However, we will need a constant $a$ that is equal to $c^2$, integrated over all variables except $r_{12}$, multiplied by $N(N-1)/2$, the 
 number of electron pairs (as they are indistinguishable).
Let us thus write one of the operators as 
\begin{equation}
 \mathfrak{W\lm} = \sum_{i<j} \mathfrak{w}(|\bfr_i - \bfr_j|,\lambda,\mu)
\end{equation}
They should be non-negligible only in the region where the approximation~\eqref{psi-asy} is valid. 
We use Eq.~\eqref{psi-asy}, and write
\begin{equation}
 \bra{\Psi\lm} \mathfrak{W\lm} \ket{\Psi\lm} \approx a \varpi(\lambda,\mu,\mathfrak{w})
 \label{frakW}
\end{equation}
where
\begin{equation}
 \varpi(\lambda,\mu,\mathfrak{w}) = \int_0^\infty d r_{12} \, 4 \pi \, r_{12}^2 \, \phi(r_{12},\lambda,\mu)^2  \, 
                                           \mathfrak{w}(r_{12},\lambda,\mu)
 \label{varpi}
\end{equation}
For example, for large enough $\mu$, one can choose $\varpi$ for  $\mathfrak{w}(r,\lambda,\mu)= \bar{w}(r,\mu)$, Eq.~\eqref{w-bar}, 
 independent of $\lambda$, or $\mathfrak{w}(r,\lambda,\mu)=(1-\lambda)  \partial_\mu w(r,\mu)$,
\begin{equation}
 \label{w'}
 \partial_\mu w(r,\mu) = \frac{2}{\sqrt{\pi}} e^{-\mu^2 r^2}
\end{equation}
For these operators, $\varpi$, the integral in Eq.~\eqref{varpi} can be obtained analytically to order $\mu^{-3}$,
\begin{widetext}
\begin{equation}
 \label{varpi-lambda} 
 \begin{split}
  \varpi(\lambda,\mu, \bar{w})  = \pi \left( \frac{1}{\mu^{2}} 
                               + \frac{\frac{4}{3 \sqrt{\pi}}\left( 1 +  2 \left( \sqrt{2}-1 \right) (1-\lambda) \right)}{\mu^{3}}   \right) \\
                               + \left[ \frac{1.53249 - 1.27957 \lambda + 0.336137 \lambda^2}{\mu^4} \right]  + \dots 
 \end{split}
\end{equation}
\begin{equation}
 \begin{split}
 \varpi(\lambda,\mu, \partial_\mu w)  = 2 \pi \, (1-\lambda) \left( \frac{1}{\mu^{3}} 
                               + \frac{\frac{2}{\sqrt{\pi}} \left( 1 +  \left( \sqrt{2}-1 \right) (1-\lambda) \right)}{\mu^{4}}  \right)  \\
                               + \left[\frac{(1-\lambda)(4.16887 - 2.19615 \lambda + 0.383474 \lambda^2) }{\mu^5} \right] +\dots  
 \end{split}
  \label{varpi-mu}
\end{equation} 
\end{widetext}
In the equations above, the terms in square brackets are uncertain, as there may be contributions to order $\mu^{-4}$ coming for 
 higher order corrections to the differential equation used to obtain $\phi$.
 
The operators $\bar{w}(r,\mu)$ and $(1-\lambda)  \partial_\mu w(r,\mu)$ were chosen because
\begin{align}
 \partial_\lambda E\lm &  = \bra{\Psi\lm} \partial_\lambda H\lm \ket{\Psi\lm} \nonumber \\
                       & = \bra{\Psi\lm}\bar{W}(\mu) \ket{\Psi\lm}      \label{E'-lambda} \\
 \partial_\mu E\lm     & = \bra{\Psi\lm} \partial_\mu H\lm \ket{\Psi\lm}  \nonumber \\ 
                       & = \bra{\Psi\lm} (1-\lambda) \partial_\mu W(\mu) \ket{\Psi\lm} \label{E'-mu}
\end{align}
To obtain $a$ we use Eq.~\eqref{frakW}, and equate the expectation values of $\mathfrak{W}$ 
 in the model system (or, equivalently the energy derivatives above)  with the asymptotic estimates using $\varpi$, 
 Eq.~\eqref{varpi-lambda}, or \eqref{varpi-mu},
\begin{align}
 \partial_\lambda E(\lambda,\mu) 
                                   & \approx a \, \varpi \left( \lambda,\mu, \bar{w} \right) \label{Ep-lambda-approx} \\
 \partial_\mu E(\lambda,\mu) 
                                 & \approx (1-\lambda) a \,  \varpi \left(\lambda,\mu,  \partial_\mu w \right)\label{Ep-mu-approx} 
\end{align}

Obtaining $a$ requires a supplementary computational effort beyond the model calculation: the expectation values in Eq.~\eqref{E'-lambda} or \eqref{E'-mu} are needed.
Using derivatives is still relatively cheap, as the model wave 
 function can be used.
The additional information provided by them is known to work for correcting models (see, e.g., Ref. \onlinecite{GutSav-PRA-07}).

\subsection{Constructing corrections}

\subsubsection*{Adiabatic connection and first-order correction}
Equalities like
\begin{align}
 E & = E(\lambda_0,\mu_0) + \int_{\lambda_0}^1  d\lambda \;  \partial_\lambda E(\lambda, \mu_0)   \label{adiab-lambda} \\
   & = E(\lambda_0,\mu_0) + \int_{\mu_0}^\infty  d\mu \;  \partial_\mu E(\lambda_0, \mu)  \label{adiab-mu}
\end{align}
are known under the name of adiabatic connections.
It seems that they were first used in quantum mechanics in the differential form by G\"uttinger, a student of Pauli.~\cite{Gut-31, Jon-15} 

Of course, even if we have solved the Schr\"odinger equation at $(\lambda_0,\mu_0)$, we do not have information about 
 $ E(\lambda > \lambda_0,\mu > \mu_0)$ needed in the equations above: the formulas seem useless.
However, for $\mu \rightarrow \infty$, the integrands appearing on the right-hand side of Eqs.~\eqref{adiab-lambda} and \eqref{adiab-mu}
 can be expressed using Eqs.~\eqref{Ep-lambda-approx} and \eqref{Ep-mu-approx}.
\begin{align}
 E & \approx E(\lambda=0, \mu_0) + a \int_0^1 d\lambda \, \varpi \left( \lambda,\mu_0, \bar{w} \right) 
        \label{E-order-1-lambda} \\
  & \approx E(\lambda=0, \mu_0) + a \int_{\mu_0}^\infty d\mu \, \varpi \left( \lambda=0,\mu,\partial_\mu w \right) 
        \label{E-order-1-mu}
\end{align}

The integrals over $\varpi$ given in Eq.~\eqref{E-order-1-lambda} or \eqref{E-order-1-mu} can be obtained analytically to order $\mu^{-3}$.
The path of integration should not matter, the two integrals in Eqs.~\eqref{E-order-1-lambda} and \eqref{E-order-1-mu} are 
 indeed identical to the order considered,
\begin{align}
 \int_0^1 d\lambda \, \varpi(\lambda,\mu_0,\bar{w}(\mu_0)) & =  \int_{\mu_0}^\infty d\mu \, \varpi(\lambda=0,\mu,\partial_\mu w) \nonumber \\
                                                       & =  \pi \left( \mu_0^{-2} + \kappa \, \mu_0^{-3}   \right) +\dots
 \label{intvarpi}
\end{align}
 where $\kappa=4 \sqrt{2}/(3 \sqrt{\pi})$.
 Let us mention that the existence of the term in $\mu^{-2}$ can be derived from simple scaling considerations, 
  while that in $\mu^{-3}$ shows up when studying how the cusp condition for the wave function is approached when
 $\mu \rightarrow \infty$.~\cite{GorSav-PRA-06}
 
For $a$, we use Eqs.~\eqref{Ep-lambda-approx} or \eqref{Ep-mu-approx} for a given value of $\lambda$ and $\mu$, e.g., $\lambda=0$ and $\mu=\mu_0$,
 as it asymptotically does not depend on $\lambda$ or $\mu$.
If we use Eq.~\eqref{Ep-lambda-approx}, we get (at $\lambda=0$),
\begin{equation}
 E \approx E(\mu) + \frac{\mu+\kappa}{\mu+\kappa(2-1/\sqrt{2})}  \,  \partial_\lambda E\lm |_{\lambda=0}  
\label{O-lambda-mu-3} 
\end{equation}
while with Eq.~\eqref{Ep-mu-approx} we obtain
\begin{equation}
 E \approx E(\mu) + \frac{\mu+\kappa}{2\mu+3\kappa} \, \mu \, E'(\mu)
\label{O-mu-3} 
\end{equation}
When a function depends on a single variable we use primes to denote the derivatives.

We can study what happens when we ignore the term that takes into account how the cusp in the wave function is approached with increasing $\mu$,
 i.e., set $\kappa$ to 0 in the formulas above.
Eqs.~\eqref{E'-lambda} and \eqref{O-lambda-mu-3} give the conventional first-order perturbation expression,
\begin{equation}
 E \approx E(\mu) + \partial_\lambda E\lm |_{\lambda=0} = \bra{\Psi(\mu)} H \ket{\Psi(\mu)} 
\label{O-lambda-mu-2} 
\end{equation}
Eq.~\eqref{O-mu-3} gives
\begin{equation}
 E \approx E(\mu) + \frac{1}{2} \mu \, E'(\mu) 
\label{O-mu-2}
\end{equation}

We see that in the limit $\mu = 0$, there is no correction with approximation ~\eqref{O-mu-2}, or ~\eqref{O-mu-3},
 so that approximation~\eqref{O-lambda-mu-2}, or \eqref{O-lambda-mu-3}, is expected to give better results.

To underline the difference between the conventional first-order perturbation theory, Eq.~\eqref{O-lambda-mu-2}, 
 and that of Eq.~\eqref{O-lambda-mu-3},
 let us add and subtract $\partial_\lambda E\lm|_{\lambda=0}$ on the right-hand side of  of the latter, and use Eq.~\eqref{O-lambda-mu-2}.
We get a term not present in standard first-order perturbation theory~\eqref{O-lambda-mu-2},
\begin{widetext}
 \begin{equation}
 E \approx  \bra{\Psi(\mu)} H \ket{\Psi(\mu)} - \frac{(1-1/\sqrt{2})\kappa}{\mu+ (2-1/\sqrt{2})\kappa } \partial_\lambda E\lm |_{\lambda=0}
 \label{h-correct} 
\end{equation}
\end{widetext}
Note that the same additional information is needed to correct $E(\mu)$ both here, and in conventional perturbation theory, namely
 $\partial_\lambda E\lm |_{\lambda=0} = \bra{\Psi(\mu)} \bar{W}(\mu) \ket{\Psi(\mu)}$.

\subsubsection*{The next step}
Let us summarize how we have obtained the approximations in Eqs.~\eqref{O-lambda-mu-3} and \eqref{O-mu-3}.
We approximated $E$ by
\begin{equation}
 E \approx E\lm + a_1 \, \chi_1\lm
 \label{in-basis-1}
\end{equation}
where $\chi_1\lm$ was chosen using the asymptotic behavior, cf. Eq.~\eqref{intvarpi}.
For Eqs.~\eqref{O-lambda-mu-2} and \eqref{O-mu-2}, we used instead $\mu^{-2}$. 
The constant $a_1 = a \pi$ was chosen by reproducing the derivative of the energy with respect to either $\lambda$ (at $\lambda=0$),
 or with respect to $\mu$, cf. Eqs.~\eqref{Ep-lambda-approx},\eqref{Ep-mu-approx}.
We take the derivative of Eq.~\eqref{in-basis-1} with respect to $\mu$, at some $\mu_0$.
In order to simplify notation, we drop here the subscript $0$.
For it, we recall that the energy of the Coulomb system does not depend on $\mu$, and get
\begin{equation}
 0 \approx E'(\mu) + a_1 \, \chi_1'(\mu)
\end{equation}
This gives $a_1 \approx - E'(\mu)/\chi_1'(\mu)$ that is substituted back in Eq.~\eqref{in-basis-1} .
If we choose $\chi_1(\mu)=\mu^{-2}$, we obtain Eq.~\eqref{O-mu-2}, while with $\chi_1(\mu)=\mu^{-2}+ \kappa \mu^{-3}$, cf. Eq.~\eqref{intvarpi},  
 we obtain Eq.~\eqref{O-mu-3}.

Let us generalize it, by writing an asymptotic series (see, e.g., Eq.~1.5.2 in Ref. ~\onlinecite{Bru-81}) for $\bar{E}$, Eq.~\eqref{E-bar},
\begin{equation}
 \bar{E}\lm \approx  S(\lambda, \mu, K) =  \sum_{k=1}^K a_k \, \chi_k\lm
 \label{in-basis}
\end{equation}
We have $\chi_k$ that vanish as $\mu \rightarrow \infty$, as $\bar{E}(\mu)$ vanishes in this limit.
Furthermore, we choose the $\chi_k$ such that
\begin{equation}
 \chi_{k+1}\lm / \chi_k\lm \rightarrow 0 \, \, \, \mathrm{for} \,  \mu \rightarrow \infty 
 \label{asymp-o}
\end{equation}
and assume that $A_K$, a positive constant, can be found, such that the truncation error,
\begin{equation}
 | \bar{E}\lm - S(\lambda,\mu,K) | \le A_K | \chi_{K+1} \lm | \, \, \, \mathrm{for} \,  \mu \ge \mu_K
 \label{asym-O}
\end{equation}
By Eq.~\eqref{in-basis}, for sufficiently large $\mu$, the truncation error is smaller for $S(\lambda,\mu,K+1)$ than for $S(\lambda,\mu,K)$.
Note, however, that for a given, small enough $\mu$, the error can increase when going from $K$ to $K+1$.
To give a simple example, approximating $(1+x)^{-1}$ by $(x^{-1}-x^{-2})$ has a smaller absolute error than $x^{-1}$ for $x>1$, 
 but the opposite applies for $x<1$.

If we had $\bar{E}\lm$, we could determine 
\begin{equation}
 a_k = \lim_{\mu \rightarrow \infty} \left( \bar{E}\lm - S(\lambda,\mu,k-1) \right)/\chi_k \lm
\end{equation}
by applying Eq.~\eqref{asymp-o}.
As we don't know $\bar{E}\lm$, we follow in this paper the generalization of the Taylor series~\cite{Wid-28}, i.e., 
 we define the $a_k$ by neglecting the terms beyond $K$, and fix the $a_k$ such that the first $K$ derivatives of $\bar{E}\lm$
 are equal to those of $S(\lambda,\mu,K)$.

A few remarks about this approach.
\begin{enumerate}[i]
 \item The derivatives of $\bar{E}\lm$ are accessible, as they are minus those of $E\lm$, as can be seen by taking
   the derivatives with respect to $\lambda$ or $\mu$ in Eq~\eqref{E-bar}.
  By changing the model, we have to change the correction in such a way that the physical energy, $E$, is unchanged.
 \item There is a similitude with conventional perturbation theory where the basis functions are $\lambda^k$, 
  and the coefficients are, up to a factor $1/k!$, the derivatives with respect to $\lambda$
  in $\lambda=0$.
 \item There is a further analogy with conventional perturbation theory: information around the expansion point (here $\lambda,\mu$, there
 $\lambda=0$) is used to get information for a system at another point (here $\mu=\infty$, there $\lambda=1$).
 \item We see that this approach is nothing but an extrapolation procedure, that can be extended, by using information in a number
  of points (the value of the function, here $E\lm$, and possibly its derivatives).~\cite{Sav-JCP-11}
 As first derivatives are relatively easy to obtain, one can also use the energies for 
  different models, and their first derivatives, and reduce by a factor two the number of models used.
 
 When derivatives don't exist, we can still use this procedure.
 For example, when we approach the Coulomb interaction by a model that does not continuously change with $\mu$ 
  (in contrast to that used here, Eq.~\eqref{w}).
 Such a model appears, for example, when we expand $1/r_{12}$ in a finite basis.
 
 The generalized Taylor expansion is a particular case: the function and the derivatives are taken in a single point.
 
 The numerical integration of the adiabatic connection formula, 
  Eq.~\eqref{adiab-lambda}, or Eq.~\eqref{adiab-mu} is another particular case that uses the first derivative of the $E\lm$ in 
   a set of points along the integration path.
\end{enumerate}

\subsubsection*{Higher order estimates}

In order to improve on the approximations already given, consider the case $K=2$.
Let it first be: $\chi_1(\mu)=\mu^{-2}, \, \chi_2(\mu)=\mu^{-3}$.
It satisfies condition~\eqref{asymp-o}.
Let us ignore for the moment the knowledge of the cusp-related relationship between $a_2$ and $a_1$, $a_2/a_1=\kappa$, cf. Eq.~\eqref{intvarpi}.
We obtain an approximation for $E$ from Eq.~\eqref{in-basis}, also taking the derivatives with respect to $\mu$, $E(\mu), E'(\mu), E''(\mu)$,
 a system of three equations with three unknowns: $E$, $a_1$, $a_2$.
After solving it, we get:
\begin{equation}
 E \approx E(\mu) +  \mu \, E'(\mu) + \frac{1}{6} \, \mu^2 E''(\mu) 
\label{O-mu-3-s} 
\end{equation}

However, as we know the relationship, we can make progress by choosing $\chi_1(\mu)=\mu^{-2}+\kappa \mu^{-3}$, 
 and $\chi_2(\mu)=\mu^{-4}$. 
By requesting the first two derivatives with respect to $\mu$ to satisfy the approximation~\eqref{in-basis}, we obtain,
\begin{equation}
 E \approx E(\mu) + \frac{7 \mu+ 4 \kappa}{8 \mu + 6 \kappa} \, \mu \, E'(\mu) + \frac{2 \mu + \kappa}{16 \mu+ 12 \kappa} \, \mu^2 E''(\mu) 
\label{O-mu-4} 
\end{equation}

Similar considerations can be applied to the expectation value of the Hamiltonian.
The term added to the expectation value of $H$ in Eq.~\eqref{h-correct}, originates from $a$ times
\begin{equation}
    - \varpi \left( \lambda=0,\mu,\bar{w}(r_{12},\mu) \right)+ \int_0^1 d\lambda \, \varpi  \lambda,\mu, \bar{w}(r_{12},\mu) \propto \mu^{-3}
\end{equation}
that can be considered as a basis function.
Its prefactor $\alpha$ can be determined by taking the derivative of
\begin{equation}
 E = \bra{\Psi(\mu)} H \ket{\Psi(\mu)} + \alpha \, \mu^{-3} + \dots
\end{equation}
with respect to $\mu$, yielding
\begin{equation}
 E \approx \bra{\Psi(\mu)} H \ket{\Psi(\mu)} + \frac{1}{3} \mu  \frac{d}{d\mu} \bra{\Psi(\mu)} H \ket{\Psi(\mu)}
 \label{h-order-2}
\end{equation}
that also contains a second-order correction, as we need the change in the wave function with $\mu$.

As adding basis functions to the expansion, Eq.~\eqref{in-basis}, are expected to improve asymptotically the approximation,
 we can use the absolute value of the difference between the estimate of the $E$ in two successive approximations, like Eq.~\eqref{O-mu-4} and 
 Eq.~\eqref{O-mu-3} as an error indicator for large $\mu$.
 
Equations like Eq.~\eqref{O-lambda-mu-3}, or \eqref{O-mu-3}, appear from the asymptotic conditions.
For sufficiently large $\mu$, fits like Eq.~\eqref{h-order-2}, or \eqref{O-mu-3-s} should yield the same result.
If this is not the case, we have an indicator that the asymptotic regime was left.
For example, as we know the relationship between the coefficients of $\mu^{-2}$ and that of $\mu^{-3}$, cf. Eq.~\eqref{intvarpi}, $a_3 = \kappa a_2$,
 we see that the added term when passing from $K=1$ to $K=2$ is smaller than the first correction, as long as $\mu > \kappa \approx 1.06$. 
It gives us a first estimate of the range of validity of the asymptotic approximation.

\subsubsection*{Asymptotic lower bounds}

An asymptotic lower bound can be constructed using the asymptotic behavior of the wave function.
We start with a trivial lower bound
\begin{align}
 E & = \frac{\bra{\Psi} H \ket{\Psi\lm}} {\bra{\Psi} \Psi\lm \rangle}  \nonumber \\
               & = E(\mu) + \frac{\bra{\Psi} \bar{W}\lm \ket{\Psi\lm}}{\bra{\Psi} \Psi\lm \rangle} \nonumber \\
               & \ge E(\mu) + \bra{\Psi} \bar{W}\lm \ket{\Psi\lm} 
\end{align}
To obtain the equalities, we used that, for all $\lambda$ and $\mu$, also for the Coulomb system, $\Psi\lm$ is an eigenfunction of $H\lm$, 
 and that $H=H\lm+\bar{W}\lm$.
For the inequality, we assumed that $\Psi\lm$ is normalized to one, for all $\lambda$ and $\mu$, 
 and that the same sign has been chosen for $\Psi\lm$ and $\Psi$.
(We could choose absolute values to avoid this choice.)
Asymptotically, we use $c \phi$, Eq.~\eqref{psi-asy}, to compute 
\begin{widetext}
\begin{equation}
     \bra{\Psi} \bar{W}(\mu) \ket{\Psi(\mu)} \rightarrow 
      a \int_0^\infty du \; 4 \pi u^2 \;\phi(u,\lambda=1,\mu=\infty) \, \bar{w}(u,\mu) \, \phi(u, \lambda ,\mu) + \dots
\end{equation}
\end{widetext}
The integral appearing on the right-hand can be evaluated analytically to order $\mu^{-3}$, to yield as expression for the lower bound to the 
 exact energy,
\begin{widetext}
 \begin{equation}
 E_{\le} \approx E\lm + a \, \pi \,  \left( \mu^{-2} 
                               + \mu^{-3} \frac{4}{3 \sqrt{\pi}} \left( 1 +  \left( \sqrt{2}-1 \right) (1-\lambda) \right)  \right)
 \label{lower-bound} 
 \end{equation}
\end{widetext}
In particular, for $\lambda=0$, we obtain Eq.~\eqref{intvarpi}, that was used to obtain the approximations in Eq.~\eqref{O-lambda-mu-3}
 and ~\eqref{O-mu-3}.
These first-order expressions are asymptotically lower bounds, while the common first-order expression $\bra{\Psi(\mu)} H \ket{\Psi(\mu)}$
is an upper bound to $E$.
This gives us further error estimates that become asymptotically error bounds.

\subsection{Varia}

\subsubsection*{Excited states and properties}
It is worth underscoring that the considerations above are also valid for excited states, as never was it assumed that $E(\mu)$ is the ground
 state energy.

Furthermore, properties can be treated the same way.
This can be done by modifying the Hamiltonian: $H \rightarrow H + \alpha A$, where $A$ is the operator of interest, and
 approaching (within the model) $\partial_\alpha E(\alpha)$ at $\alpha=0$.
 
One can also compute $A(\mu)=\bra{\Psi(\mu)} A \ket{\Psi(\mu)}$.
Remarking that the perturbation starts with $\mu^{-2}$, one concludes that $A(\mu \rightarrow \infty)$ is approached by $A(\mu)$ at most
 by terms in $\mu^{-2}$.
We can thus obtain a formula similar to that leading to Eq.~\eqref{O-mu-2}:
\begin{equation}
 A(\mu=\infty) \approx A(\mu) + \frac{1}{2} \mu \, A'(\mu)
 \label{property}
\end{equation}

\subsubsection*{Choosing the model and size-consistency}
All the considerations above are based upon the asymptotic regime.
One can perform tests, check for the stability of the results.
However, one  does not know a priori when this is reached.
If $\mu$ is chosen too small, the asymptotic approximation can fail.
If $\mu$ is chosen too large, the cost of the calculation can become unjustified.
One may compare this situation with that of choosing the right basis set: one wants it small, but not too small.

Checking for stability, or for reaching the asymptotic regime requires a new calculation.
One would like to choose the parameter on the safe side without doing one.

One can impose some supplementary condition, and this can be very useful see, e.g., Refs. \onlinecite{KroSteRefBae-JCTC-12, DalJohBec-JCP-17}.

The value of $\mu$ achieving the best compromise between cost and accuracy certainly depends on the system.
One can understand it by looking at the expression of the interaction, Eq.~\eqref{w} where $\mu$ appears only multiplying $r_{12}$.
The range separation parameter $\mu$ can be seen as a scaling parameter.
Let us consider that in a diffuse systems we are satisfied $\mu_{diffuse}$, in a compact system, we expect it that $\mu_{compact} > \mu_{diffuse}$
 has a similar effect, as $\mu$ has the dimensions of an inverse distance.
The different parts of the system may be different atoms, but also regions within an atom.
For example, the density is significantly larger in the core than in the valence region.
(In practice, this may not be very severe: the regions of high density are the core regions of the atoms,
 and for most applications these are not important, as being replaced by pseudopotentials, or giving
 contributions that are small in differences.)

Even if a prescription is found for choosing $\mu$ in a given system, there is a remaining problem.
If we decided for an optimal value for system $\mathcal{A}$, $\mu_{opt}(\mathcal{A})$, 
 and another one for system $\mathcal{B}$, $\mu_{opt}(\mathcal{B})$, this would not be a prescription
 for the system made of the two at infinite separation $\mathcal{A..B}$.
So, although all the approximations presented above guarantees size-consistency, because $E(\mu), E'(\mu), \dots$ are assumed to be accurate and thus
 size-consistent, the optimal value for $\mu$ is not prescribed.
Let us assume $\mu_{opt}(\mathcal{A}) > \mu_{opt}(\mathcal{B})$:
 if we choose $\mu_{opt}(\mathcal{A})$ the effort is too important for the sub-system $\mathcal{B}$, 
 if we choose $\mu_{opt}(\mathcal{B})$ the accuracy is not sufficient for the sub-system $\mathcal{A}$.
 
A way to get around this problem is used in DFAs: one makes a local ansatz, as it allows to decompose expression
 into contributions from the spatial regions of the subsystems, $\Omega_\mathcal{A}$ and $\Omega_\mathcal{B}$: 
\begin{equation}
  \int_{\R^3} d\bfr  \, f(\bfr)= \int_{\Omega_\mathcal{A}} d\bfr  \, f(\bfr)+ \int_{\Omega_\mathcal{B}} d\bfr  \, f(\bfr)
  \label{locality}
\end{equation}
In this paper, for the sake of simplicity, no local ansatz is made.
A dependence on the position $\bfr_1$ can be introduced into Eq.~\eqref{psi-asy} by not integrating over $\bfr_1$,
  $c \rightarrow c(\bfr_1)$.

One can choose to define local range separation parameter, based using the density in $\bfr_1 = \bfr_2$, e.g., Ref. \onlinecite{PolSavLeiSto-JCP-02}
 complemented by its gradient in this point, e.g., Refs. \onlinecite{HenIzmScuSav-JCP-07,AscKum-JCP-19},  or the local kinetic energy, e.g., 
 Fig.~3.6, p.134, in Ref. \onlinecite{Tsu-book-14}. 
This improves the results (see, e.g., Ref. \onlinecite{KruScuPerSav-08}).
However, the two-electron integrals are more complicated if the interaction is made not only to depend on the distance between
 electrons, but on their positions.
There are techniques to deal with this problem in an efficient way (see, e.g., Ref. \onlinecite{KlaBah-JCTC-20}).
However, this requires extra programming and testing.

Size-consistency issues are complicated in approximations because even in forms like that in Eq.~\eqref{locality} we have to deal with ensembles
 in the case of degeneracies. 
This is true for densities~\cite{PerParLevBal-PRL-82, Per-85, Sav-08}, and for the methods where only the short-range part of the pair density
 is needed (cf. Eq. (19) in Ref. \onlinecite{Sav-08}).
However, in general, dealing with pair densities is more complicated.
To give a simple example, consider the pair density, 
\begin{equation}
 \label{pair-density}
 P_2 (\bfr,\bfr')  = \bra{\Psi} \sum_{i \ne j}^N \delta(\bfr_i - \bfr) \delta(\bfr_j-\bfr') \ket{\Psi}
\end{equation}
Its normalization (obtained integrating it over $\bfr$ and $\bfr'$) is $N(N-1)$, because
\begin{equation}
 \int_{\R^3} d\bfr \int_{\R^3} d\bfr'  \delta(\bfr_i - \bfr) \delta(\bfr_j-\bfr') = 1
\end{equation}
 and is thus not extensive,
Nevertheless, this is not the case when one computes the expectation value of the short-range operators, 
 for example, the system average of $P_2$,
\begin{align}
 P_{2,aver} & = \bra{\Psi} \sum_{i \ne j}^N \delta(\bfr_i - \bfr_j) \ket{\Psi} \nonumber \\
            & =  \int_{\R^3} d\bfr \int_{\R^3} d\bfr' \, P_2(\bfr,\bfr') \delta(\bfr-\bfr') \nonumber \\
            & =  \int_{\R^3} d\bfr  \,\, P_2(\bfr,\bfr) 
  \label{on-top}
\end{align}
It can be decomposed into contributions from the spatial regions of the subsystems, $\Omega_\mathcal{A}$ and $\Omega_\mathcal{B}$,
 as in Eq.~\eqref{locality}, with $f(\bfr)=P_2(\bfr,\bfr)$.
 
For a detailed discussion, closely related to this issues, see Ref. \onlinecite{KroKum-20}.

\subsubsection*{Connections to DFAs}
The basic philosophy of constructing DFAs is related to that presented here
 (see, e.g., Refs. \onlinecite{ZieRauBae-TCA-77, StoGolPre-TCA-80, Bec-IJQC-83, ErnPer-JCP-98, PerRuzSunNepKap-20}).
One starts with the adiabatic connection, Eq.~\eqref{adiab-lambda} or 
 \eqref{adiab-mu}.~\cite{HarJon-JPF-74, LanPer-SSC-75, GunLun-PRB-76, Yan-JCP-98}$^,$\footnote{This approach is also related to
 that described in Ref. \onlinecite{LeeBae-PRA-95}.}
In the next step, a model is chosen for $a \, \phi^2(r_{12},\lambda,\mu)$, as in Eqs.~\eqref{Ep-lambda-approx} or \eqref{Ep-mu-approx} (however
 chosen to depend on the position $\bfr_1$).
This produces local (or semi-local) approximations, i.e., depending on $\bfr_1$.
For example, in the local density approximation, one  implicitly transfers $a \phi^2$ from the uniform electron gas 
 with a density equal to that of the system of interest in point $\bfr_1$.

To generate the models, one can use exact properties of density functionals, see, e.g., Ref. \onlinecite{Lev-PRA-91}.
One also makes an ansatz, with parameters that are obtained from calculations or from experimental data.
This parametrization can bring in new properties, desirable or not.
On one hand, such a procedure might introduce favorable features beyond those covered by the universal short-range $\phi$, Eq.~\eqref{psi-asy}. 
On the other hand, it may transfer unwanted properties, e. g., that the long-range behavior is different in the system from
 which the transfers made, e.g., in the uniform electron gas and in the system where it is used, e.g., a molecular system.
(For an analytic treatment, see, e.g., Refs. \onlinecite{LanPer-PRB-77, GilAdaPop-MP-96, TouColSav-PRA-04}.)
For example, the exchange energy starts both for an atom and the uniform electron gas with a term linear in $\mu$. 
However, the next term is cubic for the former, and quadratic for the latter.

In order to connect with DFAs, a simplified method is mentioned here.
In it, only $a$ is assumed to depend on $\bfr_1$, and the universal short-range behavior of $\phi$ is used.
One can relate $a$ to the exact system-averaged on-top pair density of the system, $P_{2,aver}$, Eq~.\eqref{on-top}.~\cite{GorSav-PRA-06}
In analogy to the local density approximation, the on-top pair density is not calculated, but approximated by that of the uniform
 electron gas having the same density as that of the system of interest in $\bfr_1$, $\rho(\bfr_1) $. 
The integration over $\bfr_1$ is performed next to yield an approximation for the ``system averaged on-top pair density''
 that is used to evaluate $a$,
\begin{equation}
 a \approx \int_{\R^3} d\bfr_1 a_{UEG} \left( \rho(\bfr_1) \right)
 \label{c2-ueg}
\end{equation}
Numerical calculations for $P_2(\bfr,\bfr)$ in uniform electron gas exist, 
 even for models such that the one used here (cf. Ref. \onlinecite{GorSav-PRA-06}).

\subsubsection*{Correcting approximations obtained by transfer}
When approximations are constructed by transfer from some accurately treated training system, e.g., the uniform electron gas,
 also asymptotic properties are inherited.
Consider that we have some approximation in this category, that allows a cheap estimate of the correction,
 $\bar{\mathcal{E}}(\mu) \approx \bar{E}(\mu)$.
If we apply the asymptotic considerations to it,  we obtain estimates for the corrections,  as above.
For example, the difference between an equation similar to Eq.~\eqref{O-mu-2}, and  Eq.~\eqref{O-mu-2}, we obtain
\begin{equation}
 \bar{E}(\mu) \approx \bar{\mathcal{E}}(\mu) + \frac{1}{2} \mu \left( E'(\mu) - \mathcal{E}'(\mu) \right)
\end{equation}
This line is not pursued in this paper.
Results obtained with this kind of approximation can be found in \cite{Sav-JCP-14}. 

\subsubsection*{Changing the one-particle potential}
As already mentioned, using the bare external potential for $V$ is not expected to work well.
As an alternative, one can explore the external potential of DFT, that ensures that the density obtained for the model is equal to the exact one.
Such calculations \cite{PolColLeiStoWerSav-IJQC-03, TouColSav-MP-05} are rigorous, can provide guidelines, 
 but are too expensive in practice.
A simplified form of is used in the numerical part of this paper, in order to explore the effect of changing the external potential.

\subsubsection*{Models for associating basis set cutoff to long-range operators} 

Many programs do not to take full advantage of model interactions like the one used in this paper.
However, programs use basis sets, and there is an analogy between using these and the range 
 separation~\cite{GinPraFerAssSavTou-JCP-18, RieKalKre-JCP-20}.
To characterize these, one needs to replace $\mu$ by parameters characterizing the range of functions, and the discretization within this range.
What makes a rigorous proof more difficult is that basis sets (e.g., Gaussian ones) are optimized to be small, and 
 are thus more difficult to characterize.
Even if a range is defined for a basis (such as by a minimal and maximal exponent, a cutoff in a plane wave basis), the 
 discretization in this range has to be controlled.
Instead, we will discuss now ways to roughly characterize a basis set using range separation.
Better (and cheaper) ways are certainly possible.

\begin{figure}[htb]
 \begin{center}
   \includegraphics[width=0.95\textwidth]{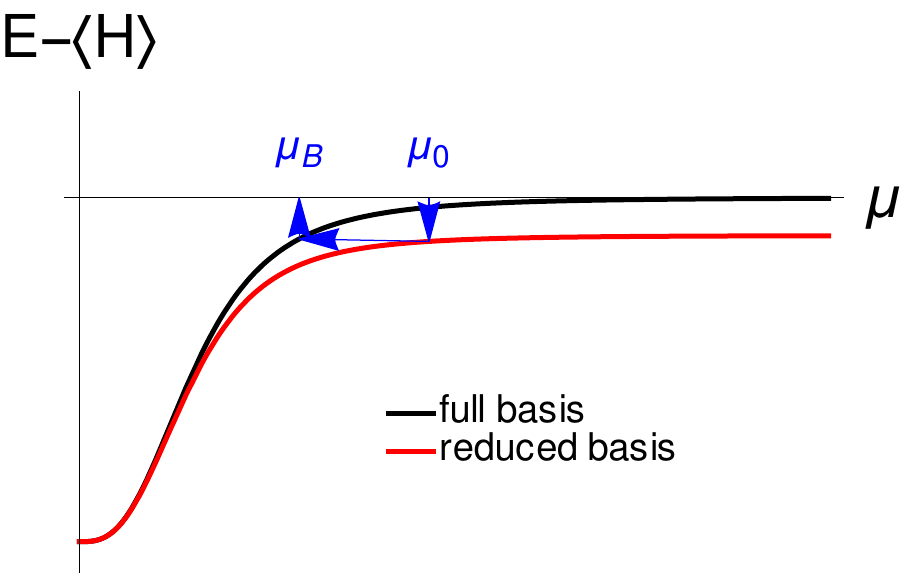}
 \end{center}
 \caption{Finding the value of the range separation parameter $\mu$ that corrects for basis set limitations.
  The correction to the expectation value of the model Hamiltonian $H(\mu_0)$ in a reduced basis set is more important than that
  in the full basis. 
  It corresponds to the correction to the expectation value of the model Hamiltonian $H(\mu_B)$ in the full basis set calculation 
  for a different model defined by $\mu_B < \mu_0$.
  }
 \label{fig:ideal-mu}
\end{figure}
Let us first discuss an ideal construction.
 and compare the expectation value of $H$ in the full basis set, $\bra{\Psi(\mu)} H \ket{\Psi(\mu)}$
 with that obtained with a wave function $\Psi_B$ optimized in a reduced basis set $B$, $\bra{\Psi_B(\mu)} H \ket{\Psi_B(\mu)}$. 
For the non-interacting system, usually even ``small'' basis sets are good enough.
This remains true, as long as the interaction is weak.
However, we are interested in applying asymptotic corrections, and we have to continue to stronger interactions.
As this happens,  $\bra{\Psi_B(\mu)} H \ket{\Psi_B(\mu)}$ is 
 expected to be above  $\bra{\Psi(\mu)} H \ket{\Psi(\mu)}$.
Thus, the correction needed to obtain $E$, $E-\bra{\Psi_B(\mu)} H \ket{\Psi_B(\mu)}$, is more important than that needed to correct
 $\bra{\Psi(\mu)} H \ket{\Psi(\mu)}$, cf. Fig.~\ref{fig:ideal-mu}.
We note that for a given model, defined by $\mu_0$ we can find a $\mu_B$, such that the corrections are the same,
\begin{equation}
  E-\bra{\Psi_B(\mu_0)} H \ket{\Psi_B(\mu_0)} = E- \bra{\Psi(\mu_B)} H \ket{\Psi(\mu_B)}
  \label{ideal-mu-B}
\end{equation}
If the calculation in the reduced basis set was performed at $\mu_0$, the correction should be that obtained for $\mu_B \le \mu_0$ in the 
 full basis set.
In short: in order to correct $\bra{\Psi_B(\mu_0)} H \ket{\Psi_B(\mu_0)}$, the correction at $\mu_B$ should be used, not that at $\mu_0$.

\begin{figure}[htb]
 \begin{center}
   \includegraphics[width=0.95\textwidth]{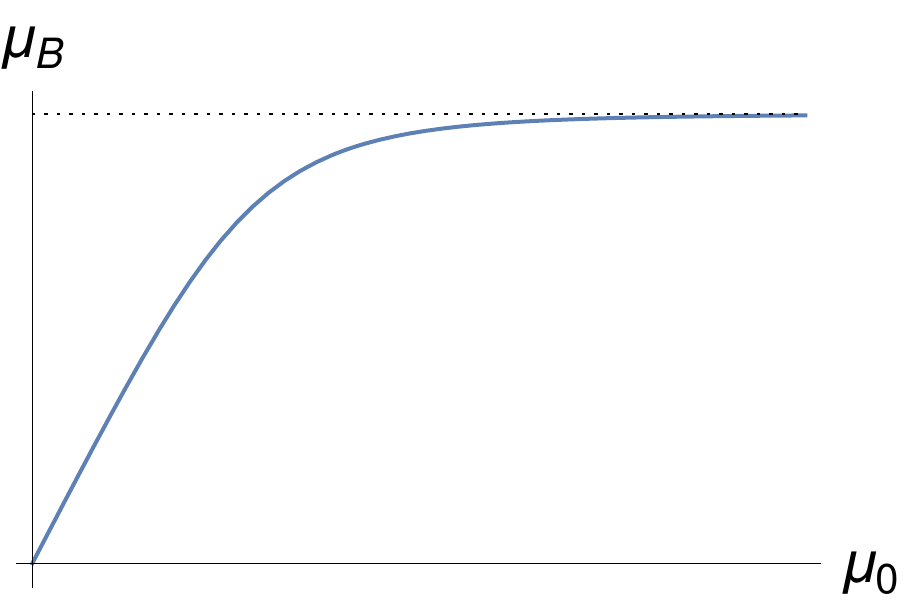}
 \end{center}
 \caption{The model parameter $\mu_B$ to be used for the correction of $\bra{\Psi_B(\mu_0)} H \ket{\Psi_B(\mu_0)}$.
          The horizontal dotted line corresponds to the value obtained in the reduced basis set $B$ with the Coulomb
          potential and corrects only for the basis set error.
  }
 \label{fig:mub-of-mu0}
\end{figure}
The dependence of $\mu_B$ on $\mu_0$ is schematically shown in Fig.~\ref{fig:mub-of-mu0}.
Note that as $\mu_0 \rightarrow \infty$, $\mu_B$ approaches a finite value: even 
 the lowest energy expectation value obtained in the basis $B$ for the Hamiltonian with Coulomb interaction,
 $ \min_{\Psi_B} \bra{\Psi_B} H \ket{\Psi_B}$,  has to be corrected. 

Basis sets (such as Gaussian) have different quality in different regions of space:
 their ability to describe the interaction varies from one part of the system to another.
This suggests to use local corrections of the basis set deficiency, cf. Eq.~\eqref{locality}.
Although this is not presented in this paper, it can be found in the literature.~\cite{GinPraFerAssSavTou-JCP-18}

Basis set correction solves two problems in range-separation.
First, it solves the problem of choosing an optimal value for the range-separation parameter.
One can choose the parameter $\mu$ large, even stay with the Coulomb interaction,
 and the basis set introduces automatically a change of the model parameter to $\mu_B$.
Second, locality is introduced through the use of a local basis set, and no adjustment of the two-electron integrals is needed;
 only in the corrections locality has to be taken care of.

The prescription defining $\mu_B$ as above,  Fig.~\ref{fig:ideal-mu}, is not useful in practice, as it uses the expectation value of $H$ 
 in the full basis.
To better understand the origin of the difference between the calculations in the full and in the reduced basis set,
 let us first introduce the projector on the basis set,
\begin{equation}
 \mathcal{P}_B  = \sum_{i,j \in B} \ket{\Phi_i} ( S^{-1} )_{i,j} \bra{\Phi_j}
 \label{proj-B}
\end{equation}
 where $\Phi_i$ are the $N$-particle basis functions, and $S^{-1}$ is the inverse of the matrix of their overlap.
We note that in a given basis set, expectation values obtained with the operator
$\mathcal{P}_B W(\mu) \mathcal{P}_B$ are the same as those obtained with $W(\mu)$.
Of course, this is also true for other operators, such as $\bar{W}$, or $W'(\mu)$.

Note that $\mathcal{P}_B W \mathcal{P}_B$ acting on a basis function produces a linear combination of the basis functions.
It does not describe the strong increase in the interaction when two electrons approach; 
 it is unable to describe the singularity of the Coulomb potential.
In the reduced basis set, the electrons do not interact via the Coulomb potential.

\begin{figure}[htb]
 \begin{center}
   \includegraphics[width=0.95\textwidth]{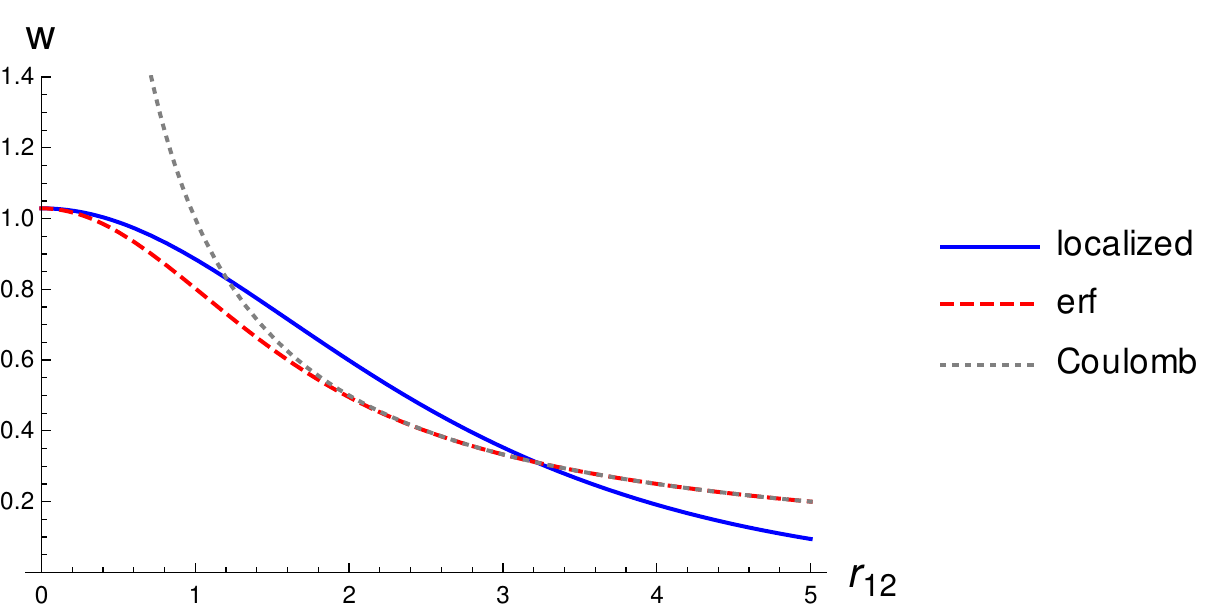}
 \end{center}
 \caption{Interaction potentials as function of the inter-particle distance.
  Coulomb potential: dotted, black curve.
  Model potential $w$, Eq.~\eqref{w}: red, dashed curve.
  Localized potential $w_{loc}$, generated from the Coulomb potential projected onto a finite basis, 
   satisfying Eq.~\eqref{mu-B-loc}: blue, full curve.
  }
 \label{fig:w-loc}
\end{figure}

In order to compare the non-local interaction generated by $\mathcal{P}_B W \mathcal{P}_B$ with the model interaction $w(\mu)$, Eq.~\eqref{w}, 
 let us first obtain a local interaction potential, $w_{loc}$.
There are many ways of doing it.
Here, we will use one close in spirit to Slater's local potential.
To have $\bra{\Psi_B(\mu)} W_{loc}(\mu) \ket{\Psi_B(\mu)}$ equal to 
 $\bra{\Psi_B(\mu)} \mathcal{P}_B W \mathcal{P}_B \ket{\Psi_B(\mu)}$ before 
 we integrate over the distance between electrons, and write
\begin{widetext}
\begin{equation}
  \bra{\Psi_B(\mu)}  \delta(|\bfr_1-\bfr_2|-u) w_{loc}(|\bfr_1-\bfr_2|,\mu) \ket{\Psi_B(\mu)} = 
  \bra{\Psi_B(\mu)}  \delta(|\bfr_1-\bfr_2|-u) \mathcal{P}_B W \mathcal{P}_B \ket{\Psi_B(\mu)}
\end{equation} 
\end{widetext}
For the sake of simplicity, the formula is given for the system with two electrons, but can be generalized to more electrons.
As we have expanded $\Psi_B$ in a basis,
\begin{equation}
 \Psi_B(\mu) = \sum_k b_k(\mu) \Phi_k
 \label{psi-in-B}
\end{equation}
and $w_{loc}$ depends only on the distance between electrons over which we do not integrate, we obtain, in matrix form,
\begin{equation}
 w_{loc}(u,\mu) = \frac{\mathbb{B}^\dagger(\mu) \mathbb{D}(u) \mathbb{W}(\mu) \mathbb{B}(\mu)}
                      {\mathbb{B}^\dagger(\mu) \mathbb{D}(u) \mathbb{C}(\mu)} 
\label{w-loc}
\end{equation}
Here, $\mathbb{B}$ is the vector of coefficients $b_k$, Eq.~\eqref{psi-in-B},
      $\mathbb{D}(u)$ is the matrix with elements $\bra{\Phi_i}   \delta(|\bfr_i-\bfr_j|-u) \ket{\Phi_j}$, and
      $\mathbb{W}(\mu)$ is the matrix with elements $\bra{\Phi_i}   w(|\bfr_i-\bfr_j|,\mu) \ket{\Phi_j}$
To choose $\mu_B$, we find a model potential, Eq.~\eqref{w}, that resembles $w_{loc}$ from
\begin{equation}
 w(r_{12}=0, \mu_B) = w_{loc}(r_{12}=0,\mu)
 \label{mu-B-loc}
\end{equation}
as sketched in Fig.~\ref{fig:w-loc}.
Its behavior is dictated by the functions present in the projection operator.
We see that $w_{loc}$ is finite at the origin, and decays to rapidly at large distances (where the localized basis set does not reach).
The latter shortcoming is not so important in practice, as the pair density vanishes anyhow in that region.

For the asymptotic corrections, we further need the coefficient $a$, that we can generate as in Eq~\eqref{frakW}, and
 assuming from the similarity of the operators that we can use $\Psi_B(\mu) \approx \Psi(\mu_B)$.
\begin{widetext}
\begin{equation}
 \bra{\Psi_B(\mu)} \bar{W}(\mu_B) \ket{\Psi_B(\mu)} \approx\bra{\Psi(\mu_B)} \bar{W}(\mu_B) \ket{\Psi(\mu_B)} 
                                               \approx a \, \varpi(\lambda=0,\mu_B, \bar{w} (r_{12},\lambda=0,\mu_B)))
 \label{c2-muB}
\end{equation}
\end{widetext}
This requires the knowledge of $\bra{\Psi_B(\mu)} \bar{W}(\mu_B) \ket{\Psi_B(\mu)}$.
If we wish to avoid calculating this new integral, we can exploit the freedom of choosing the operator $\mathfrak{W}$ for determining $a$,
 Eq.~\eqref{frakW} and \eqref{varpi}, and choose a different $\mu$ for $\Psi$ and for $\mathfrak{W}$,
\begin{widetext}
\begin{equation}
 \bra{\Psi_B(\mu)} \bar{W}(\mu) \ket{\Psi_B(\mu)} \approx a \int_0^\infty d r_{12} \, 4 \pi \, r_{12}^2 \, \phi(r_{12},\lambda=0,\mu_B)^2  \, 
                                        \bar{w} (r_{12},\lambda=0,\mu)
 \label{c2-mixed} 
\end{equation} 
\end{widetext}
With this expression, $\bra{\Psi_B(\mu)} \bar{W}(\mu) \ket{\Psi_B(\mu)}$ is needed, a quantity already used for obtaining the 
 $\bra{\Psi_B(\mu)} H \ket{\Psi_B(\mu)}$.

\section{Numerical results}

\subsection{Producing numerical results}

\subsubsection*{System: Harmonium}
Numerical results will only be presented for Harmonium.
Its Hamiltonian has the external one-particle potential
\begin{equation}
 \label{v-harm}
 v(r)=\frac{1}{2} \omega^2 r^2
\end{equation}
For $\omega=1/2$, and a pair of electrons ($N=2$) the non-interacting ($\mu=0$) exact wave function is given by 
 a product of harmonic oscillator eigenfunctions, 
\begin{equation}
 \label{psi-harm}
 \psi(\bfr)= (\omega/\pi)^{3/4} e^{-\frac{1}{2} \omega r^2}
\end{equation}
For the Coulomb interaction the exact ground state wave function is also known (see, e.g. Refs. \onlinecite{Tau-93, KarSze-10}):
\begin{equation}
 \label{Psi-harm}
 \Psi(\bfr_1,\bfr_2,\mu=\infty)= \psi(\bfr_1) \psi(\bfr_2) (1+\frac{1}{2}|\bfr_1-\bfr_2|)
\end{equation}
times the singlet spin function.
The exact energy is known, $E(\mu=\infty)=2$.
For analyzing energy differences one might think of the ionization potential.
However, for our method where $V$ is the external potential, the one-electron system is treated exactly.
Thus, the errors in the ionization potential equal those of the total energy.
However, the errors can be different in the ground and the excited state.
The excited states can be obtained accurately.~\cite{KarCyr-CMST-03,KarCyr-04}
One excited state is treated here, the first of the same symmetry as the ground state,  $E(\mu = \infty)=2.9401169\dots$.~\cite{Kin-96}

To treat the case of $\mu \in (0,\infty)$,
we use a basis set, inspired by the Hylleraas ansatz~\cite{Hyl-30, Kin-96}:
\begin{equation}
 \Psi(\bfr_1,\bfr_2) =  \psi(\bfr_1) \psi(\bfr_2) \sum_{i,j,k} c_{ijk} s^i t^j u^k
 \label{ansatz-Hyl}
\end{equation}
where $s=r_1+r_2$, $t=r_2-r_1$, and , as above, $u=|\bfr_1-\bfr_2|$.
The powers $i$,$j$,$k$ used are the same nine as used by Hylleraas in his treatment of the Helium atom.~\cite{Hyl-30}
We note that this basis is able to reproduce the limiting cases, $\mu=0$, and $\mu=\infty$.
A comparison with a calculation on a grid (in the style of Ref. \onlinecite{GonAyeKarSav-TCA-16}) shows that the error with the 
 this basis set is maximally of 0.1 mhartree, for all $\mu$.
In order to study a basis sets effect, a reduced basis set was also used, obtained by eliminating the terms having odd powers in $u$.
The retention of even powers of $u$ allows the description of angular correlation, as $u^2=r_1^2+r_2^2-2 \bfr_1\cdot\bfr_2$.
However, omitting the even powers of $r=r_{12}$ produces an energy error of $\approx 7$~mhartree for $\mu=\infty$, even
 if the error is lower by an order of magnitude than that produced by ignoring the interaction (at $\mu=0$).
In order to underline the importance of the term linear in $u$, let us mention that retaining only the terms $i=j=0$, $k=0$ and $1$
 in Eq.~\eqref{ansatz-Hyl} is capable of reproducing the energy with an error of maximally 1~mhartree, for all $\mu$.
 
The calculations were done with Mathematica~\cite{Wol-20}, except for the $\mu$-dependent local density approximation, 
 $\mu$-LDA~\cite{Sav-INC-96, PazMorGorBac-PRB-06} energies that
 were produced with Molpro~\cite{Molproshort-PROG-12}.
 
Mathematica~\cite{Wol-20} was also used for the formal derivations in this paper.
 
\subsubsection*{Obtaining energy derivatives}
In order to obtain the derivatives, the quantities (the energy, or the property) were calculated on a grid of values of $\mu$,
 between 0 and 3~bohr$^{-1}$.
The data were interpolated, and the derivatives of the interpolant were used.
Explicit formulas can be used, but this was not needed because efficiency was not of concern for the simple case of Harmonium.
 
\subsubsection*{Description of the plots}
\label{sec:description}
Numerical calculations are made to explore how simple the model can 
 be made when using asymptotic corrections, i.e.,
 how small the parameter $\mu$ can be made, and still retain chemical accuracy ($\approx 1$~kcal/mol).
To this end, many plots show the error made by the model (corrected, or not).
A white background indicates the domain of chemical accuracy, while a gray background is used outside it.

Note that in this calculations $\mu=0$ corresponds to the bare external potential, Eq.~\eqref{v-harm}.
Thus, $E(\mu)$ is $0.5$~hartree below the exact energy, while
 $\bra{\Psi(\mu=0)} H \ket{\Psi(\mu=0)}$ is $\approx 0.06$~hartree above it, and $\approx 0.03$~hartree above the Hartree-Fock energy.
As this errors are very large, the domain of the plot is chosen to include the domain of the error made in variational calculation
 in the reduced basis excluding the odd powers of $u$ in the ansatz~\eqref{ansatz-Hyl}.

\subsubsection*{Smallest acceptable $\mu$}
By making the corrections asymptotically correct, we expect the errors to become smaller than chemical accuracy for 
 all $\mu$ larger than some value that we designate as the ``smallest acceptable $\mu$'' (SA$\mu$). 
 
As we know from second-order perturbation theory, with weak interactions only near-degenerate states contribute significantly 
 to the energy (see, e.g., Fig. 9 in Ref. \onlinecite{Sav-INC-96}).
Furthermore the lower $\mu$, is the better the features of long-range interaction can be exploited numerically;
 it is desirable to have small SA$\mu$s.

\subsection{Discussion of the results}

\subsubsection*{First-order corrections}
\begin{figure}[htb]
 \begin{center}
   \includegraphics[width=0.95\textwidth]{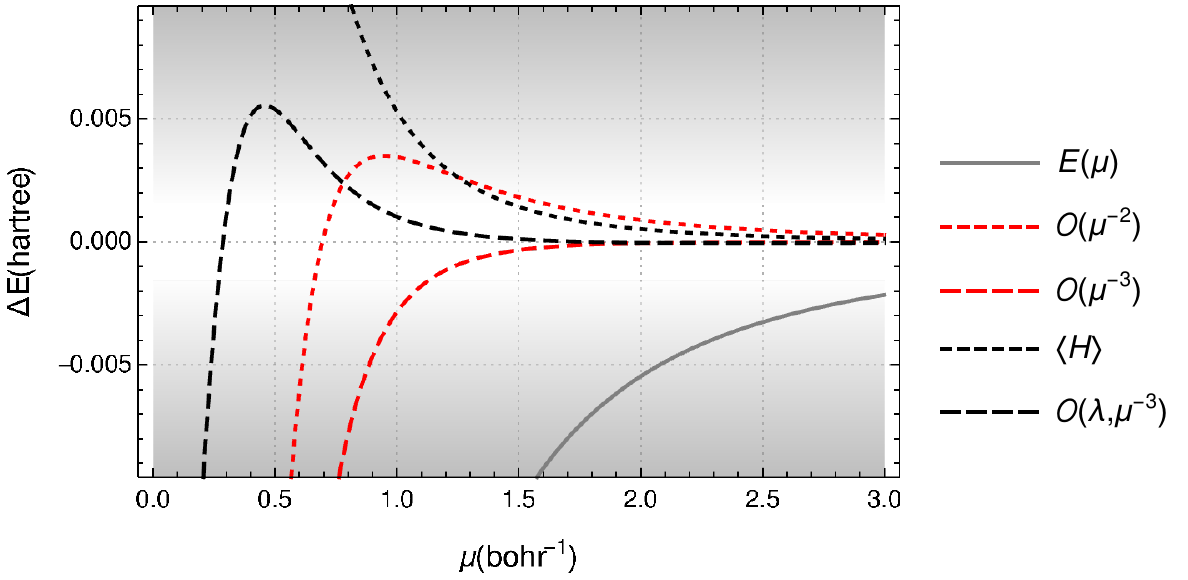}
 \end{center}
 \caption{Energy errors with first-order methods.
  Model energy, not corrected, Eq.~\eqref{SE}: full, gray curve.
  Expectation value of $H$ with the model eigenfunction: black, short-dashed curve. 
  Model corrected to order $\mu^{-2}$, Eq.~\eqref{O-mu-2}: red, short-dashed curve.
  Model corrected to order $\mu^{-3}$, Eq.~\eqref{O-mu-3}: red, long-dashed curve.
  Model corrected to order $\lambda$, taking into account terms to $\mu^{-3}$, Eq.~\eqref{O-lambda-mu-3}: black, long-dashed curve.
 \label{fig:order-1}
 }
\end{figure}
We start by looking at the model energies, $E(\mu)$, Fig.~\ref{fig:order-1}.
As the $W(\mu)$ is weaker than the Coulomb interaction, $E(\mu) < E$, for all finite $\mu$ .
The values of $E(\mu)$ are very low, and chemical accuracy is not yet reached for the largest value of $\mu$ of the domain shown 
 in the figure.
Fig.~\ref{fig:order-1} shows also the effect of the first-order corrections, obtained by using not only $E(\mu)$, but also a
 derivative of it with respect to either $\lambda$, Eq.~\eqref{O-lambda-mu-3},  or $\mu$, Eq.~\eqref{O-mu-2} or \eqref{O-mu-3}.
 
As stated above, $\bra{\Psi(\mu)} H \ket{\Psi(\mu)}$, obtained from the knowledge of  $\partial_\lambda E(\lambda,\mu)$ at $\lambda=0$,
 yields unacceptably high errors when $\mu$ is small.
However, already for a $\mu$ larger than $\approx 1.5$~bohr$^{-1}$ chemical accuracy is reached.

A result, very close to the preceding (if not slightly better, as having not too large errors, down to almost $0.5$~bohr$^{-1}$)
 is obtained using $E'(\mu)$ and taking into consideration the existence of a $\mu^{-2}$ term, Eq.~\eqref{O-mu-2}.
While the previous correction uses the expectation value of  $\bar{W}(\mu)$, the latter one uses that of $W'(\mu)$.
When taking into account the existence of the cusp, the same information can be used to include terms to order $\mu^{-3}$,  Eq~\eqref{O-mu-3}.
Asymptotically the values get better, as they should.
However, we note a deterioration at smaller $\mu$ when we do not expect the asymptotic regime to work.
In the example treated it remains a lower bound for all $\mu$.

The SA$\mu$ is lowered when the asymptotic considerations are applied to perturbations in $\lambda$, Eq.~\eqref{O-lambda-mu-3}, that
 is also an asymptotic lower bound.
It approaches the correct value faster than the other approximations, 
This is better than the traditional upper bound (obtained with the expectation value of $H$) for large $\mu$.
However, at small $\mu$, the lower bound property is lost, and becomes even worse than the upper bound for $\mu \approx 0$, not shown in the figure.

\subsubsection*{Second-order and first-order error estimates}
\begin{figure}[htb]
 \begin{center}
   \includegraphics[width=0.95\textwidth]{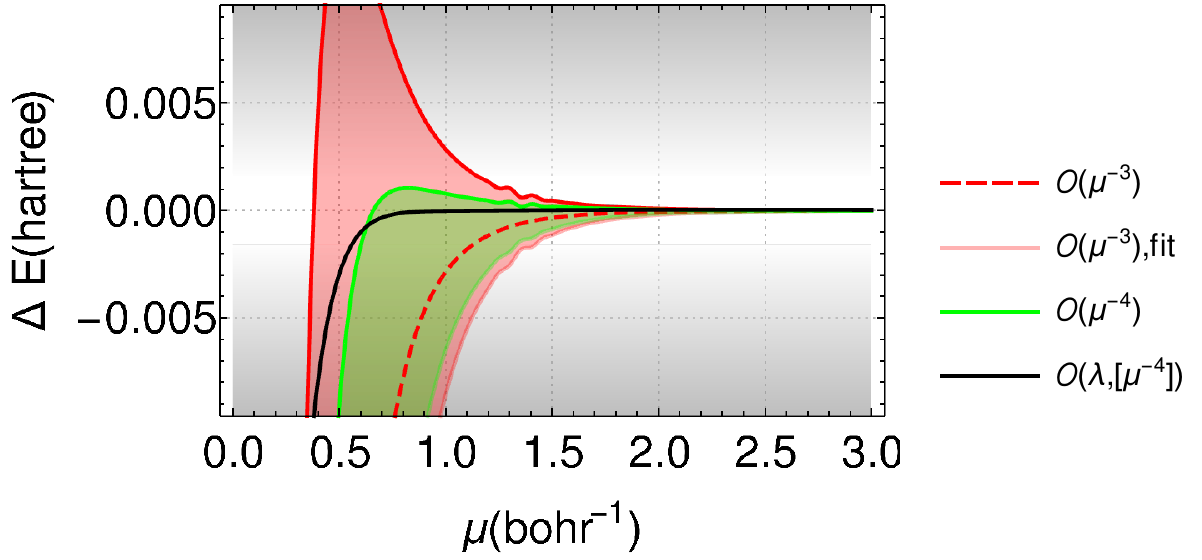}
 \end{center}
 \caption{Energy errors at second order.
  First-order correction, asymptotically correct to order $\mu^{-3}$, Eq.~\eqref{O-mu-3}: red, long-dashed curve.
  Also as a reference, first-order correction, asymptotically correct to order $\mu^{-3}$, but keeping the uncertain term in $\mu^{-4}$ 
   showing up in Eq.~\eqref{varpi-lambda}: black, full curve.
  Second-order correction, with fitted coefficient of $\mu^{-3}$ term, Eq.~\eqref{O-mu-3-s}: red, full curve.
  The difference to the first-order value, correct to $\mu^{-3}$ leads to an error estimate, shown by a pink filling.
  Second-order correction, with fitted coefficient of $\mu^{-4}$ term, Eq.~\eqref{O-mu-4}: green, full curve.
  The difference to the first-order value, correct to $\mu^{-3}$ leads to an error estimate, shown by a dark green filling.
  }
 \label{fig:order-2}
\end{figure}
To estimate the error we can use the upper and asymptotic lower bound in Fig.~\ref{fig:order-1}.
Let us now use $E''(\mu)$.
First we consider the coefficient of $\mu^{-3}$ as a parameter to fit,  instead of being given by the
 asymptotically valid relationship.
A comparison of the two approximation, Eq.~\eqref{O-mu-3-s} and Eq.~\eqref{O-mu-3}, is shown in Fig.~\ref{fig:order-2}.

Another possibility is to use $E''(\mu)$, to correct Eq.~\eqref{O-mu-3} by adding a term in $\mu^{-4}$, Eq.~\eqref{O-mu-4}.
This approximation lowers the SA$\mu$ to $\approx 0.5$~bohr$^{-1}$.

To be on the safer side,  the difference to the first-order approximation (going to $\mu^{-3}$), can be used to estimate asymptotically the error,
  see Fig~\ref{fig:order-2}.

The terms to order $\mu^{-4}$ in Eq.~\eqref{varpi-lambda} can be used in an exploratory way, without using $E''(\mu)$,
 although they are not believed to be exact.
The errors are practically zero for all $\mu$, down to almost $0.5$~bohr$^{-1}$, see Fig.~\ref{fig:order-2}.
There is yet no explanation for this good behavior.

\begin{figure}[htb]
 \begin{center}
   \includegraphics[width=0.95\textwidth]{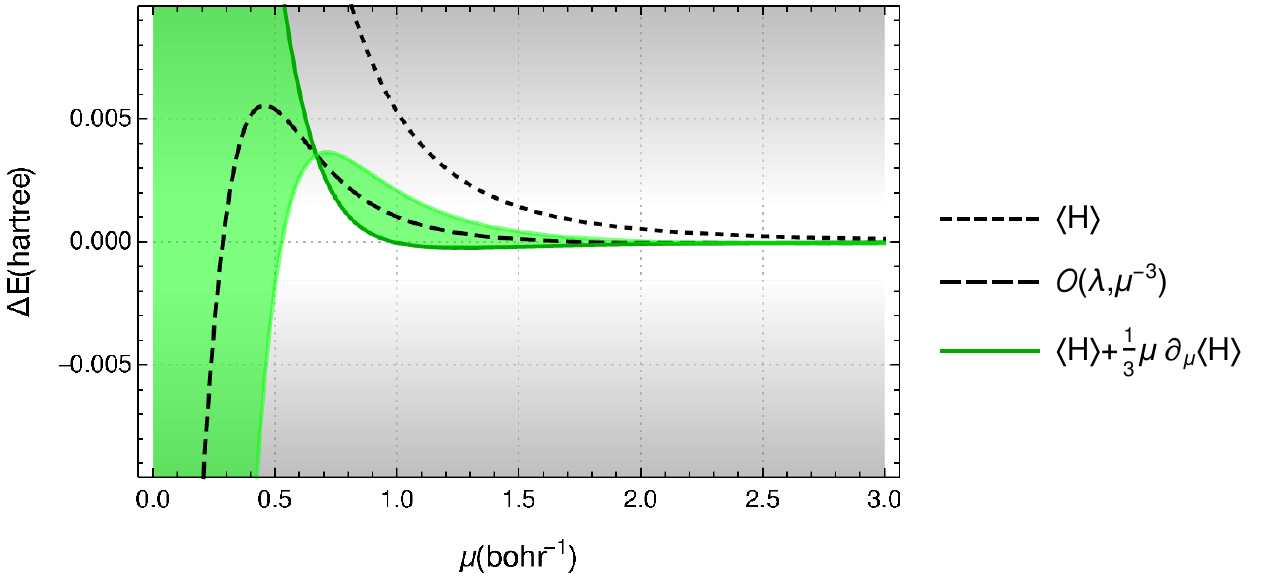}
 \end{center}
 \caption{Energy errors at second order.
  Expectation value of $H$ with the model eigenfunction: black, short-dashed curve. 
  Model corrected to order $\lambda$, taking into account terms to $\mu^{-3}$, Eq.~\eqref{O-lambda-mu-3}: black, long-dashed curve.
  Second order correction, using $\partial_\mu \bra{\Psi(\mu)} H \ket{\Psi(\mu)}$,  Eq.~\eqref{h-order-2}: green, full curve.
  The difference to first-order leads to an error estimate, shown by a light green filling.
  }
 \label{fig:h-order-2}
\end{figure}
One can also consider Eq.~\eqref{h-order-2} using $\partial_\mu \bra{\Psi(\mu)} H \ket{\Psi(\mu)}$, also of second order.
For values of $\mu$ down to $1$~bohr$^{-1}$, the errors are negligible for the present discussion, see Fig.~\ref{fig:h-order-2}.
It can be also use to estimate the asymptotic errors of the first order approximation, Eq.~\eqref{h-correct}.
Note, however, that for smaller values of $\mu$, the error estimate reaches zero (before exploding for even smaller values of $\mu$).
It happens because the asymptotic regime has been left, the lower bound property was lost.
However, the present example seems to indicate that the asymptotic lower bound expression, Eq.~\eqref{O-lambda-mu-3} still gives a better energy
 lower than the upper bound expression, Eq.~\eqref{O-lambda-mu-2} for a large range of $\mu$ (and when the errors are within chemical accuracy).

\subsubsection*{Excited state}

\begin{figure}[htb]
 \begin{center}
   \includegraphics[width=0.95\textwidth]{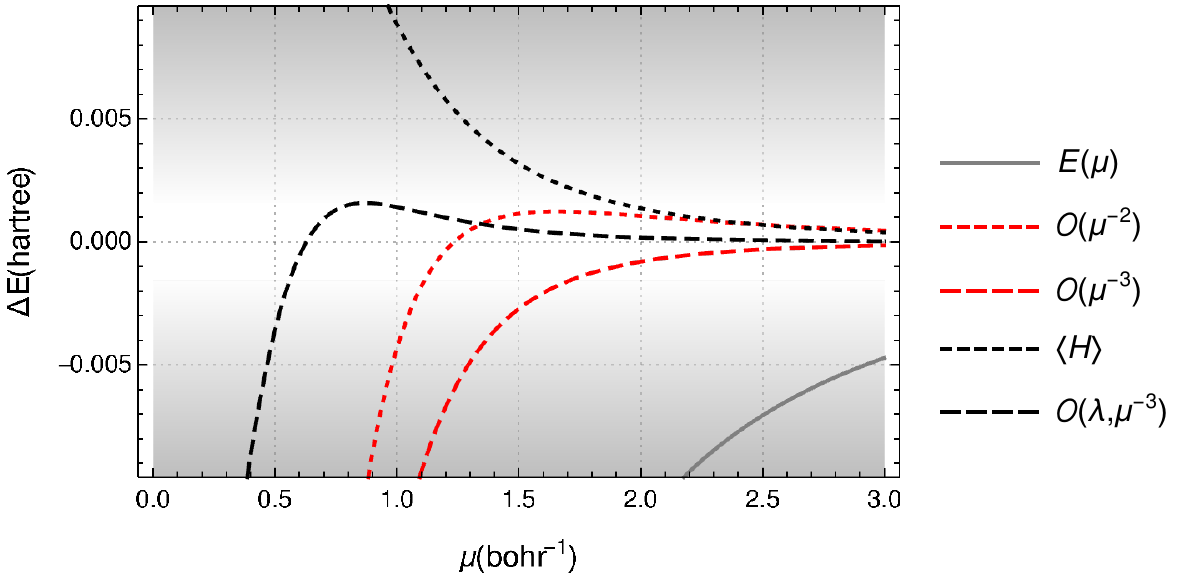}
 \end{center}
 \caption{Energy errors for first excited $^1S_g$ state.
  Model energy, not corrected, Eq.~\eqref{SE}: gray, full curve.
  Expectation value of $H$ with the model eigenfunction: black, short-dashed curve. 
  Model corrected to order $\mu^{-2}$, Eq.~\eqref{O-mu-2}: red, short-dashed curve.
  Model corrected to order $\mu^{-3}$, Eq.~\eqref{O-mu-3}: red, long-dashed curve.
  Model corrected to order $\lambda$, taking into account terms to $\mu^{-3}$, Eq.~\eqref{h-correct}: black, long-dashed curve.
  }
 \label{fig:error-excit}
\end{figure}
We now consider an excited state, the first excited $^1S_g$.
It was chosen for two reasons.
First, it is considered to have two dominant configurations, $1s 3s$ and $2p^2$, cf. Ref. \onlinecite{Kin-96}.
Second, it has the same symmetry as the ground state.
Occasionally, one finds in literature that the Hohenberg-Kohn theorem can be used for the lowest state in each symmetry.
This example shows that one can apply the asymptotic expressions also to states of the same symmetry.
The curves shown in Fig.~\ref{fig:error-excit} are obtained with the same methods as for the ground state, Fig~\ref{fig:order-1}.
The general features resemble those noticed for the ground state, and will not be repeated here.

\subsubsection*{Properties}

\begin{figure}[htb]
 \begin{center}
   \includegraphics[width=0.95\textwidth]{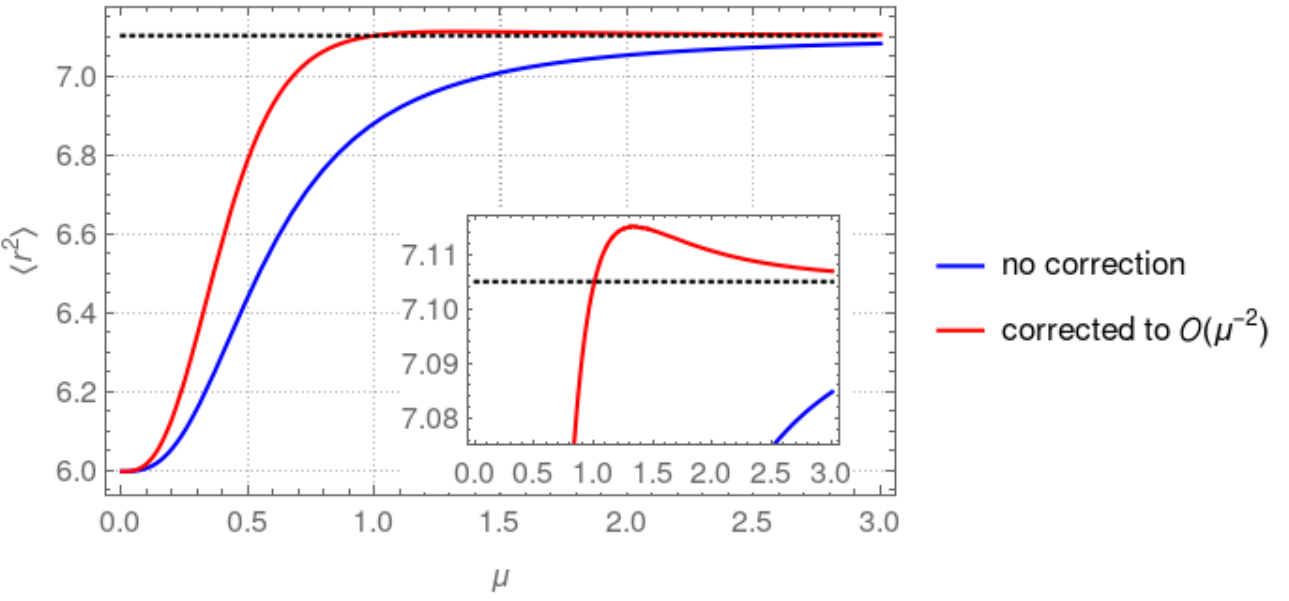}
 \end{center}
 \caption{Expectation value of $r^2$.
  Provided by the model, blue curve.
  Corrected to order  $\mu^{-2}$, red curve.
  The value for the system with Coulomb interaction is shown by a horizontal dotted line.
  The inset reduces the range of the plot, to better show the quality of the correction.
  }
 \label{fig:r2-expect}
\end{figure}

For illustrating the application to properties, let us consider a measure of the size of system, the expectation value
\begin{equation}
 \bra{\Psi(\mu)} \bfr_1^2 + \bfr_2^2 \ket{\Psi(\mu)} = \int_{\R^3} d\bfr  \, \rho(\bfr, \mu) r^2
\end{equation}
In (exact) density functional theory, it would be exact, as the density is considered exact for all models ($\mu$).
However, for approximations, this is not the case.
For the model considered here, where $V$ is the bare potential, the system is significantly more compact for small $\mu$ than for large $\mu$.

We note (see Fig.~\ref{fig:r2-expect}) that the first-order asymptotic correction improves the quality of the result for all $\mu > 0$, and
 yield errors smaller by at least an order of magnitude, for $\mu$ down to $\approx 1$~bohr$^{-1}$.

\subsubsection*{Effect of changing the one-particle potential}

\begin{figure}[htb]
 \begin{center}
   \includegraphics[width=0.95\textwidth]{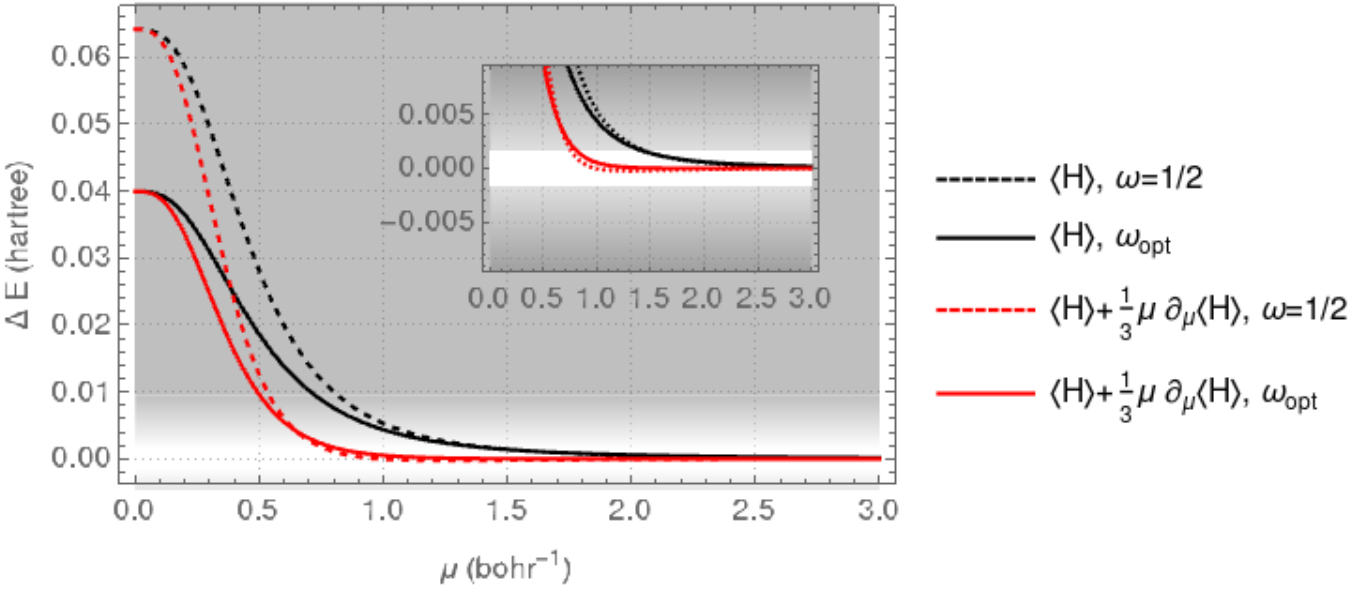}
 \end{center}
 \caption{The effect on optimizing the external potential ($\omega$) of the model on energy errors.
 The expectation value of $H$ with $\Psi(\mu)$ with the bare potential ($\omega=1/2)$: black, dashed curve.
 The expectation value of $H$ with $\Psi(\mu)$ with $\omega$ optimized for each $\mu$: black, full curve.
 The expectation value of $H$ with $\Psi(\mu)$ with second-order correction, Eq.~\ref{h-order-2}, 
  and the bare potential ($\omega=1/2)$: red, dashed curve.
 The expectation value of $H$ with $\Psi(\mu)$ with $\omega$  with second-order correction, Eq.~\ref{h-order-2}, and 
  $\omega$ optimized for each $\mu$: red, full curve.
 The inset shows the same curves in a reduced range, namely that of most plots in this paper. 
  }
 \label{fig:error-opt-omega}
\end{figure}

As it is common knowledge that a mean-field one-particle potential is better than the bare potential, we explore it 
 by a prescription that requires the knowledge of the exact density.
The purpose of this exercise is to find out how important the change of $V$ affects the asymptotic conditions.
Of course, no general conclusion can be drawn from a single example.

There are several ways to replace $V$ by a mean-field approximation, one of them being the density functional (Kohn-Sham)~\cite{KohSha-PR-65} 
 prescription to keep the density equal to the exact one for all models.
A convenient way to reach it, is to use the Legendre transform, see, e.g., Refs. \onlinecite{Lie-IJQC-83, ColSav-JCP-99},
\begin{widetext}
\begin{equation}
 v(\mu) = \arg\max_v \left( \min_\Phi \bra{\Phi} T + V + W(\mu) \ket{\Phi} - \int_{\R^3} d\bfr \, \rho(\bfr, \mu=\infty) \, v(\bfr) \right)
 \label{v-Lieb}
\end{equation}
\end{widetext}
the usual Kohn-Sham potential corresponding to $\mu=0$.

In the present	 exploration of $v$, we use a restricted maximization: we maximize not over $v$, but over those of Harmonium, Eq.~\ref{v-harm}.
Instead of taking $\omega=1/2$, we consider now $\omega$ a parameter that can be optimized at each $\mu$ to bring the density of
 the model system close to the exact one, in the sense of Eq.~\eqref{v-Lieb}.

Fig.~\ref{fig:error-opt-omega} shows the errors of the expectation values of $H$, corrected or not, as a function of $\mu$.
The optimization of the potential leads to a significant improvement at small $\mu$.
At $\mu=0$, although $v$ was only partially optimized, the expectation value of $H$ is close to the Hartree-Fock one.
As the expectation value obtained with the Kohn-Sham determinant is above the latter, we see that we must be very close to the Kohn-Sham solution.
The error is nevertheless important: the correlation energy is missing.

Let us now consider what happens when the errors reach chemical accuracy.
The values obtained with, or without optimization of the one-particle potential $v$ are very close.
In order to rationalize this surprising result, one may argue that as long as one is in the domain where
 the universal character dominates, the nature of the external potential does not play an important role.
However, please recall, that the point where the switching between the regimes occurs depends on the system: the value
 of $\mu$ for which the one-particle potential becomes important is larger in a dense system.

\subsubsection*{Comparison with a DFA}

\begin{figure}[htb]
 \begin{center}
   \includegraphics[width=0.95\textwidth]{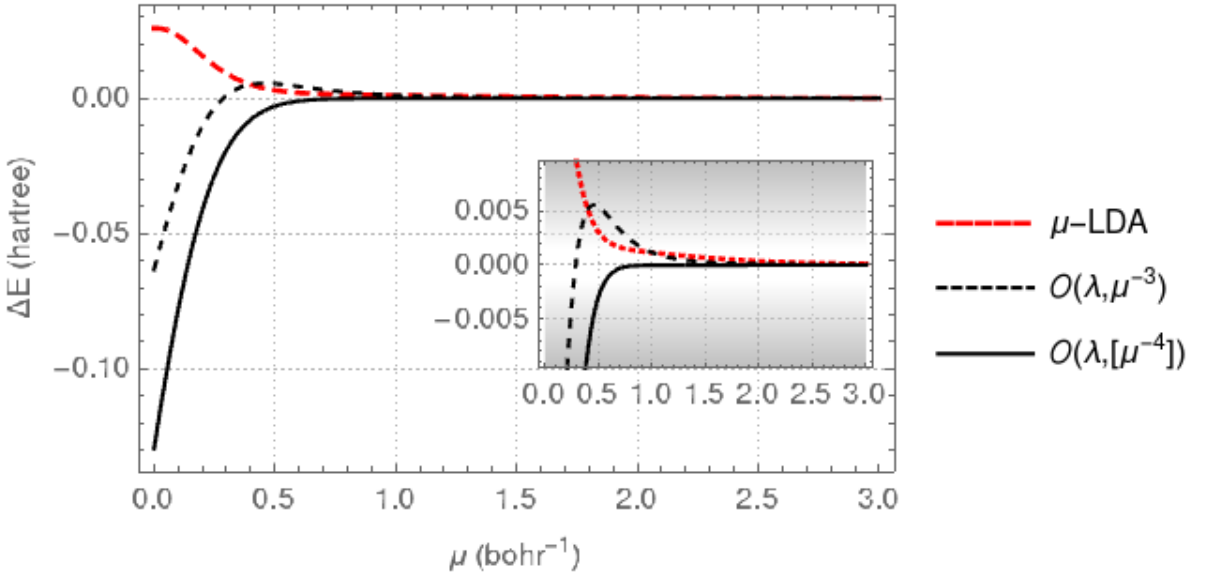}
 \end{center}
 \caption{Energy errors with $\mu$-LDA: red, long-dashed curve. To be compared to first-order approximations, 
 i) asymptotically correct to order $\mu^{-3}$, Eq.~\eqref{h-correct}, black, short-dashed curve, ii) asymptotically correct to order $\mu^{-3}$, 
   but keeping the $\mu^{-4}$ term in Eq.~\eqref{varpi-lambda}: black, full curve.
 The inset shows the same curves in a reduced range, namely that of most plots in this paper.
  }
 \label{fig:lda-error}
\end{figure}

Let us now make a single comparison with a DFA.
We choose the $\mu$-dependent local density approximation, $\mu$-LDA.
It corrects the models to obtain the exact energy of the uniform electron gas for all $\mu$
 (while the usual LDA does it only for $\mu=0$).
It is considered to be a good approximation for large $\mu$.
For example, one of the extensions to the Perdew-Burke-Ernzerhof (PBE) method to include the dependence on $\mu$, requires that it behaves like
 $\mu$-LDA for large $\mu$.~\cite{GolWerSto-PCCP-05}
We see in Fig.~\ref{fig:lda-error} that $\mu$-LDA  is indeed a good approximation for large $\mu$.
We also see that it works well in the same range as the asymptotic methods discussed in this paper.

\subsubsection*{Using the uniform electron gas to explore the validity of the asymptotic approximation.}
\begin{figure}[htb]
 \begin{center}
   \includegraphics[width=0.95\textwidth]{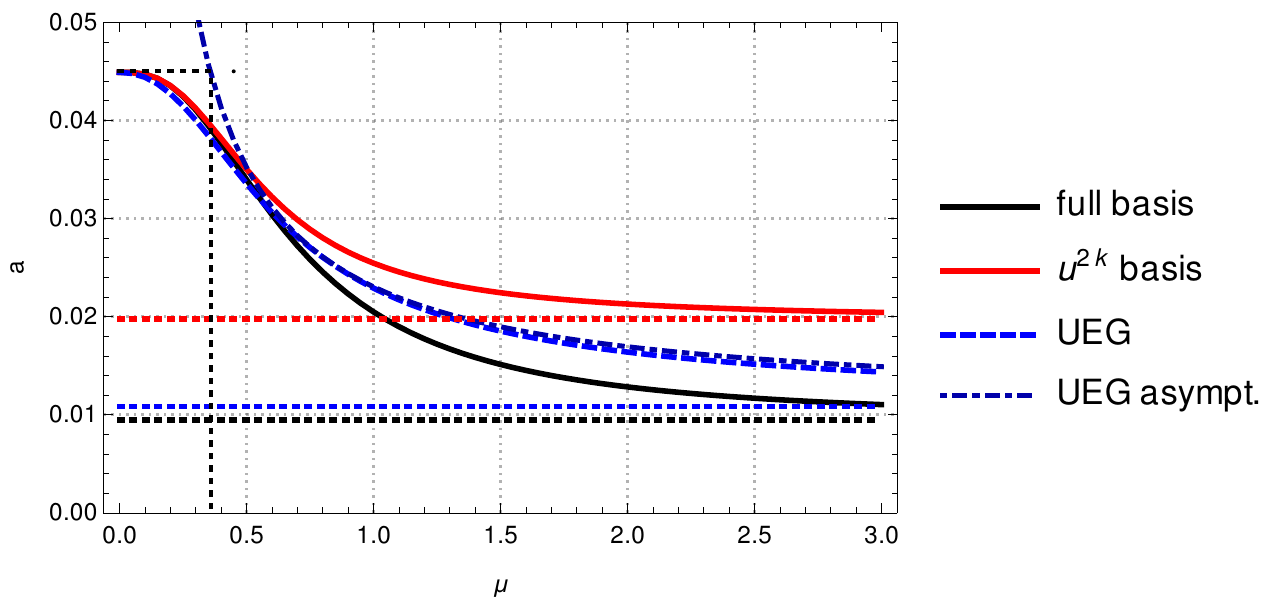}
 \end{center}
 \caption{On-top pair density system average as a function of $\mu$.
 Full basis: black, full curve.
 Reduced basis (no terms in odd powers of $r_{12}$): red, full curve.
 Based upon uniform electron gas calculations: blue, dashed curve.
 Their values at $\mu=\infty$ are indicated by horizontal lines of the same color.
 Using solely the large $\mu$ behavior of the uniform electron gas: blue, dot-dashed curve (diverging at the origin).
 Dotted lines mark the point where the asymptotic extrapolation of the uniform electron gas is equal to the value 
  of the on-top pair density of the non-interacting system, $P_{2,aver}(\mu=0)$, Eq.~\ref{on-top-mu-0}.
  }
 \label{fig:lda-c2}
\end{figure}
We now turn to the question whether using the uniform electron gas in a local approximation, Eq.~~\eqref{c2-ueg},
 can give us an easy-to-use estimator of validity of the asymptotic approximation.
 
In Fig~\ref{fig:lda-c2} we compare some accurate values of the system-averaged on-top pair density, $P_{2,aver}$, Eq.~\eqref{on-top}, known
 with different methods.
The exact value for Coulomb interaction is obtained exactly, $1/(16 \pi + 10 \pi^{3/2})$.
The uniform electron gas values used are from a model that is supposed to be accurate for short-range, the Overhauser model.~\cite{GorSav-PRA-06}

We see that as $\mu$ increases, the calculation with the large basis set yields significantly lower values for $P_{2,aver}$
 than those obtained with a basis set not containing terms in odd powers in the inter-electronic distance.
Using the electron gas expression is interesting because the density provides values that have little sensitivity to the basis set.
In Fig.~\ref{fig:lda-c2} the reduced basis set was used, but on the scale of the plot, the difference between basis sets would hardly be seen.
Fig.~\ref{fig:lda-c2} shows that the transfer from the uniform electron gas provides better results.
Asymptotically, we know~\cite{GorSav-PRA-06}
\begin{equation}
 P_{2,aver}(\mu) = \left(1+\frac{2}{\sqrt{\pi} \mu} \right) P_{2,aver}(\mu=\infty)
 \label{on-top-asy}
\end{equation}
for all systems.
When we use the uniform electron gas estimate for $P_{2,aver}(\mu=\infty)$ we get a curve that diverges at $\mu=0$.
It crosses the curve obtained using a basis set for the system of interest, that is finite at $\mu=0$.
We see that this happens at $\mu \approx 0.5$, that is the SA$\mu$ for the better of our approximations.

If we want to avoid calculations of $P_{2,aver}(\mu)$ for our system, we could compute the cheaper (and less basis-set sensitive)
\begin{align}
 P_{2,aver}(\mu=0) & = \frac{1}{2} \int_{\R^3} d\bfr \, \left( \rho(\bfr)^2 - \rho_\alpha(\bfr)^2 - \rho_\beta(\bfr)^2 \right) \nonumber \\
                   & = \int_{\R^3} d\bfr \, \rho_\alpha(\bfr) \rho_\beta(\bfr)
 \label{on-top-mu-0}
\end{align}
where the connection between the on-top pair density and spin-components of the density have been used (~\cite{YamFue-77}; 
 for the relationship to DFT, see Refs. \onlinecite{BecSavSto-TCA-95, PerSavBur-PRA-95}).
The values of $\mu$ for which  the uniform electron gas evaluated Eq.~\eqref{on-top-asy} gives a higher value should not be considered reliable.
This estimate can be seen as an a priori estimate (that can be made before the expensive calculation is performed).
However, it is too optimistic, the SA$\mu$ is too low.

\subsubsection*{Basis set errors}

\begin{figure}[htb]
 \begin{center}
   \includegraphics[width=0.95\textwidth]{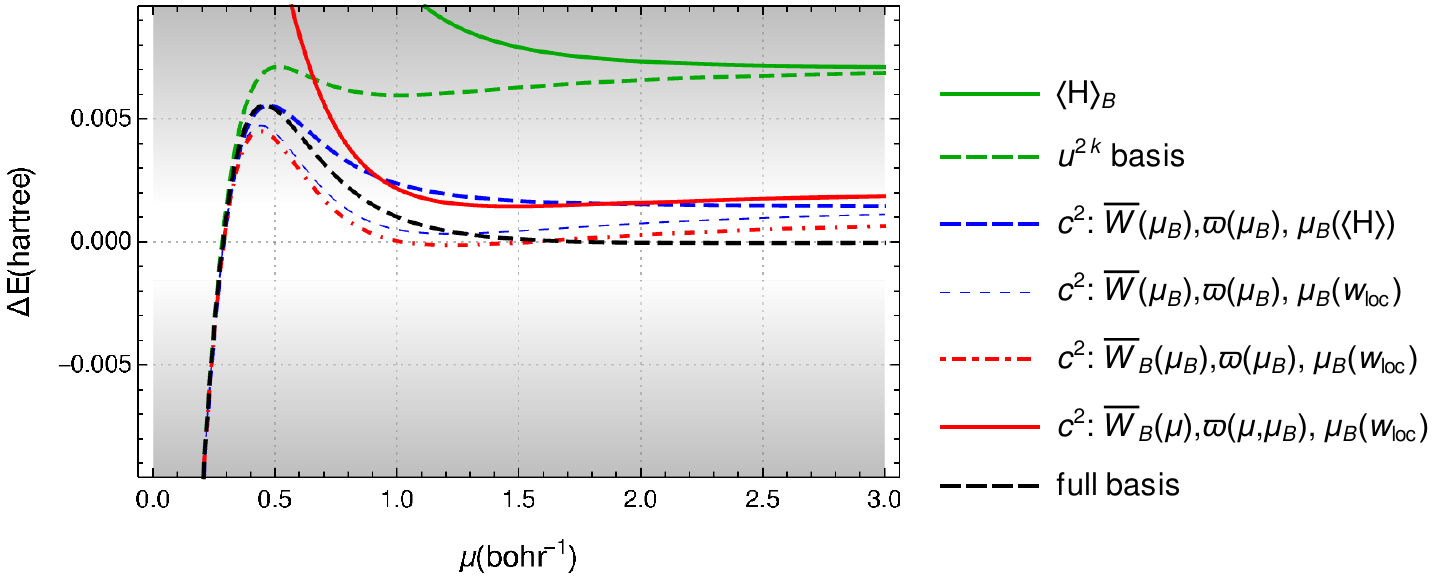}
 \end{center}
 \caption{Energy errors in a reduced basis set calculation. 
 The expectation value of the physical Hamiltonian in the reduced basis, $\bra{\Psi_B(\mu)} H \ket{\Psi_B(\mu)}$: green, full curve.
 All corrections are first-order in $\lambda$, Eq.~\ref{h-correct}.
 As a reference, the result in the full basis set: black, long-dashed curve.
 The following use all $\bra{\Psi_B(\mu)} H \ket{\Psi_B(\mu)}$, but differ in the corrections.
 Using the expectation value of $\bra{\Psi(\mu)} \bar{W}(\mu_B) \ket{\Psi(\mu)}$ to obtain the correction: blue curves;
 using the ``ideal'' $\mu_B$, Eq.~\eqref{ideal-mu-B}, Fig.~\ref{fig:ideal-mu}: thick dashed curve.
 using $\mu_B$ determined from $w_{loc}(u=0)$, Eq.~\eqref{mu-B-loc}: thin dashed curve.
 Using $\bra{\Psi_B(\mu)} \bar{W}(\mu_B) \ket{\Psi_B(\mu)}$: dot-dashed red curve.
 Using $\bra{\Psi_B(\mu)} \bar{W}(\mu) \ket{\Psi_B(\mu)}$: full red curve.
  }
 \label{fig:basis-set-error}
\end{figure}

Let us now try to correct basis set errors with range separation.
We compare the results obtained with the reliable basis set described above with one where the terms containing terms in odd powers
 of the distance between electrons are absent.
In Fig.~\ref{fig:basis-set-error}, one can see that the expectation value $\bra{\Psi_B(\mu)} H \ket{\Psi_B(\mu)}$
 has errors well beyond chemical accuracy, although it is correlated in the sense that it is able to describe radial and angular correlation.
Simply applying the asymptotic lower bound formula, Eq.~\eqref{h-correct}, works well, but it approximates
 Coulomb system in the reduced  basis,  $\bra{\Psi_B} H \ket{\Psi_B}$, not in the full basis.
One may argue that the result is better, because the error of the lower bound provided by
 the asymptotic approximation, Eq.~\eqref{h-correct} is still above the exact result, and thus better than the upper bound.
However, chemical accuracy is reached only for a a very narrow domain of $\mu$, difficult to establish.
 
Let us now correct the basis set errors using $\mu_B$. 
First, we correct $\bra{\Psi_B} H \ket{\Psi_B}$ with the ``ideal'' $\mu_B$, obtained by using the full basis expectation values of $H$,
 as in Eq.~\eqref{ideal-mu-B}, or Fig.~\ref{fig:ideal-mu}.
The correction is calculated at this $\mu_B$, using the full basis wave function, viz., $\bra{\Psi(\mu)} \bar{W}(\mu_B) \ket{\Psi(\mu)}$.
We see that the error does not vanish asymptotically.
This is because, for the reduced basis we use, $\mu_B$ is limited to a value $< 1$~bohr$^{-1}$.
We cannot get an improvement larger than the asymptotic correction gives for this value.
Next, let us successively 
\begin{enumerate}[i]
 \item change $\mu_B$ by that given by the prescription in Eq.~\eqref{mu-B-loc}, but still requesting full basis information,
 \item use $\bra{\Psi_B(\mu)} \bar{W}(\mu_B) \ket{\Psi_B(\mu)}$ with Eq.~\eqref{c2-muB}, i.e., that works only in the
       reduced basis set at $\mu$, but requires the evaluation of a new integral, and 
 \item use $\bra{\Psi_B(\mu)} \bar{W}(\mu) \ket{\Psi_B(\mu)}$ with  Eq.~\eqref{c2-mixed}, that uses only existing information.
\end{enumerate}
The different variants discussed above give all results that are within chemical accuracy for $\mu>1$.

These results suggest that once the basis set is good enough to make the asymptotic correction active, it can easily be corrected within reasonable 
 accuracy.

\section{Conclusions and perspectives}
\subsubsection*{Formal results}

In the formal part of the paper, it was shown that it is possible to build models that approach the physical, Coulomb system
 by choosing a family  of  operators that do not show a singularity as the distance between electrons reduces to zero, Eq.~\eqref{w}.
It is called range separation as the long-range is treated by the model, and the corrections have to
 deal with the missing part.

Here, the corrections are obtained by considering the solution of the Schr\"odinger equation as the distance between electrons reduces to zero, 
 Eq.~\eqref{srSE}, and not as nowadays frequently done by using density functionals.
However, the method presented here can be seen as justifying density functional approximations.
This short-range behavior is universal (independent of the external potential, i.e., that of the interaction between
 nuclei and electrons).
In this paper this is supplemented by system-specific information: the derivatives of the model energies with respect to parameters
 that describe the exact system is approached, a form of perturbation theory.
The first-order is easily accessible, as it does not need new wave functions.
Alternatively, several models (values of $\mu$) could be used to generate equations that can be solved to obtain $E$.

A maybe simpler way to describe the present approach is to realize that a model system can be corrected with an expansion in a basis set, 
 Eq.~\eqref{in-basis}.
The basis set functions are chosen according to the way the model approaches the Coulomb system.
The expansion coefficients are determined (in this paper) by using information from the model system, namely
 how its energy evolves with model parameters (the derivatives of the energy).
One can see it as an adjustment of the evolution of the model towards the asymptotic (universal) behavior.

The proposed expressions for approximating $E$ are given in Eqs.~\cref{O-mu-2,O-mu-3,O-lambda-mu-3,h-correct,O-mu-3-s,O-mu-4,h-order-2}.
For properties, similar expressions can be found, Eq.~\eqref{property}.

The corrections improve asymptotically ground state energies, 
  excited state energies,  
  and properties. 
In this regime, they can be systematically improved, in the sense that with more computational effort (more derivatives)
 the results get better, and the errors can be estimated. 
At the same time, an estimation to test if the asymptotic regime is reached is possible.

Size-consistency can be satisfied, because the formulas are linear in the model energies and their derivatives, when the model is solved accurately. 

The question arises about what happens when the models are not solved accurately.
A typical limitation is the use of a finite basis set.
Basis set errors can be modeled by associating a model interaction to the effective interaction potential. 
By using this trick, the basis set calculation can be associated to different model calculation, and the correction applied is that of 
 the latter.

\subsubsection*{Formal perspectives}

There are several formal steps that could be pursued.
In this paper, only the information from the terms up to $\mu^{-3}$ was used.
Knowledge about higher order terms exists and can be applied (for triplet electron pairs~\cite{GorSav-PRA-06}, or also for singlet 
 electron pairs, in the spirit of Refs. \onlinecite{Tew-08, RasChi-JCP-96c}).
This knowledge could get a firmer basis by a mathematical treatment, as it is done for the Coulomb 
 interaction.~\cite{FouHofHofOst-CMP-09, Yse-10, FlaFlaSch-20}

The expansions considered are asymptotic expansions: the results can be made to work better as we approach the physical system,
 but they do not necessarily get better, as we get further away from it.
This regime is reached in different systems for different values of the expansion parameter, and local forms are useful to treat such cases.
Although not described in this paper, the methods presented are easily generalized to local forms, like in DFAs, cf. Eqs.~\eqref{locality} and
 \eqref{c2-ueg}.

It is not evident how to reach the non-interacting system.
A possible path might be to use basis functions $\chi_i$, Eq.~\eqref{in-basis}, that do not diverge at $\mu=0$.~\cite{Sav-JCP-11}  
 They can satisfy the known behavior for $\mu \rightarrow 0$ (see, e.g., Ref. \onlinecite{TouColSav-PRA-04}).
Density functional approximations may be a source of inspiration.~\cite{Sav-JCP-14}
 
In this paper the bare external potential is used. 
Exploring ways to improve on it was largely used in DFAs, and can be also pursued in the present context.

The numerical technique used in this paper for obtaining the energy derivatives (beyond the first) was essentially a finite difference method.
Explicit formulas should be developed for higher order corrections.

The interaction used in this paper is convenient, but certainly not the best. 
It is not necessary that the models approach in a continuous way the Coulomb system; the derivatives used in this paper can be
 replaced by a series of models~\cite{Sav-JCP-11}.
It is worth exploring the tensor product decomposition in this context.

One does not have to connect the model system with the Coulomb system by modeling the interaction.
It is possible to use non-local one-body operators, see, e.g., Refs.  \onlinecite{ReySav-IJQC-98, AyaScuSav-CPL-99, GutSav-PRA-07, TeaHelSav-16}. 

Asymptotic expansions and their extensions is an active field of mathematics (see, e.g., Ref. \onlinecite{Boy-99}).
Without doubt, the methods presented here can be improved.

\subsubsection*{Conclusions from numerical explorations}
Numerical calculations are needed to explore when the asymptotic regime is reached, and how far one has to go with the expansion.
This paper presents results for the Harmonium.
Its purpose is to illustrate the methods proposed, not to assess their quality.

The corrections improve the model results significantly, not only for the ground state, but also for excited states, and properties, 
 cf. Figs~\ref{fig:order-1}, \ref{fig:error-excit}, and \ref{fig:r2-expect}.
It seems that although the expansion parameter used is $1/\mu$, the asymptotic regime is valid for values close, and even below $\mu=1$, cf.
 Figs~\ref{fig:order-2}, \ref{fig:h-order-2}.
The formulas lead asymptotically to lower bounds, that are at least as good as the usual upper bound, the expectation
 value of $H$, with practically no additional computational effort.

Taking into account the simplicity of the model intended to deal with the basis set correction, one must remark that it works astonishingly well,
cf Fig.~\ref{fig:basis-set-error}.
 
It was surprising that the use of the bare potential was not critical, once that the corrections were activated.
One might risk to attribute it to the universality of the correction.

Applying the model to the uniform electron gas show that - as long as the errors are within chemical accuracy -
 density functional and the asymptotic correction give similar results, cf. Fig.~\ref{fig:lda-error}. 
Also, the uniform electron gas can provide a (crude) check if the asymptotic regime is reached, cf. Fig. \ref{fig:lda-c2}.

As conclusions drawn for a single system are not sufficient, they should be treated with circumspection, although results
 in the literature  are in line with the present results for Harmonium.
Many more calculations are needed to strengthen them.

\section{Acknowledgements}

The author is grateful to
Roi Baer (The Hebrew University of Jerusalem, Israel),
Evert Jan Baerends (Vrije Universiteit Amsterdam, The Netherlands),
Maria J. Esteban (CNRS and Universit\'e Paris-Dauphine, France),
Heinz-J\"urgen Flad (Technische Universit\"at M\"unchen, Germany),
Emmanuel Fromager (Strasbourg University, France), 
Paola Gori-Giorgi (Vrije Universiteit Amsterdam, The Netherlands),
Kimihiko Hirao (RIKEN, Kobe, Japan),
Jacek Karwowski (Nicolaus Copernicus University, Toru{\'n}, Poland),
Leeor Kronik (Weizmann Institute, Rehovoth, Israel),
Stephan K{\"u}mmel (University of Bayreuth, Germany),
Katarzyna Pernal (\L{}\'od\'z University of Technology, Poland),
\'Etienne Polack (Universit\'e Bourgogne Franche-Comté, France),
Gustavo E. Scuseria (Rice University, Houston, USA), 
Hermann Stoll (University of Stuttgart, Germany),  and
Henryk Witek (National Chiao Tung University, Hsinchu, Taiwan)
for reading the first draft of the manuscript and making many suggestions to improve it.

\section{Data availability statement}

The data that support the findings of this study are available upon request.

\end{document}